\documentclass[aps,prl,twocolumn,showpacs,notitlepage,floatfix,superscriptaddress]{revtex4-1}
\usepackage{amsfonts}
\usepackage{mathtools}
\usepackage{graphicx}
\usepackage{epsfig}
\usepackage{dcolumn}
\usepackage{bm}
\usepackage{amsmath}

\usepackage[normalem]{ulem}
\usepackage[utf8]{inputenc}
\usepackage{ulem}
\usepackage{epstopdf}
\usepackage{subfigure}
\usepackage{color}
\usepackage[dvipsnames]{xcolor}
\usepackage{amsthm}
\usepackage{newlfont}
\usepackage{graphicx}
\usepackage{amssymb}
\usepackage{epstopdf}
\usepackage{appendix}
\usepackage[breaklinks=true]{hyperref}
\usepackage{breakcites}
\usepackage{textcomp}
\usepackage{appendix}
\usepackage{multirow}	
\usepackage{color}
\usepackage[dvipsnames]{xcolor}
\usepackage{amssymb}
\usepackage{epsfig}
\usepackage{bm}
\usepackage[american]{babel}
\usepackage{braket}
\hypersetup{colorlinks=true,linkcolor=RubineRed,citecolor=Blue,filecolor=Blue,urlcolor=Blue,pdfstartview=FitH}
\usepackage{orcidlink}

\makeatletter
\def\maketitle{
\@author@finish
\title@column\titleblock@produce
\suppressfloats[t]}
\makeatother

\begin{document}
\title{Spontaneous Quantum Turbulence in a Newborn Bose-Einstein Condensate via the Kibble-Zurek Mechanism}

\author{Seong-Ho Shinn\orcidlink{0000-0002-2041-5292}}
\email{seongho.shin@uni.lu}
\affiliation{Department of Physics and Materials Science, University of Luxembourg, L-1511 Luxembourg, Luxembourg}

\author{Matteo~Massaro\orcidlink{0009-0004-0935-8022}}
\email{matteo.massaro@uni.lu}
\affiliation{Department of Physics and Materials Science, University  of Luxembourg, L-1511 Luxembourg, Luxembourg}

\author{Mithun Thudiyangal\orcidlink{0000-0003-4341-6439}}
\affiliation{Center for Quantum Technologies and Complex Systems (CQTCS), Christ University, Bengaluru, Karnataka 560074, India}
\affiliation{Department of Physics and Electronics, Christ University, Bengaluru, Karnataka 560029, India}

\author{Adolfo del Campo\orcidlink{0000-0003-2219-2851}}
\affiliation{Department of Physics and Materials Science, University of Luxembourg, L-1511 Luxembourg, Luxembourg}
\affiliation{Donostia International Physics Center, E-20018 San Sebasti\'an, Spain}

\begin{abstract}
The Kibble-Zurek mechanism (KZM) predicts the spontaneous formation of topological defects in a continuous phase transition driven at a finite rate. We propose the generation of spontaneous quantum turbulence (SQT) via the KZM during Bose-Einstein condensation induced by a thermal quench. 
 Using numerical simulations of the stochastic projected Gross-Pitaevskii equation in 
two spatial dimensions, we describe the formation of a newborn Bose-Einstein condensate proliferated by quantum vortices. We establish the nonequilibrium universality of SQT through the Kibble-Zurek and Kolmogorov scaling of the incompressible kinetic energy.
\end{abstract}
\maketitle

Understanding turbulence is one of the long-standing unsolved problems in physics, with broad applications in biology, medicine, industry, and weather prediction \cite{Kolmogorov41_first, Frisch1995}. Feynman pioneered the idea of quantum turbulence (QT) as an analog in superfluids proliferated by a tangle of vortices \cite{Feynman1955}. With the development of technology to create, manipulate, and probe quantum fluids, 
 the study of quantum turbulence emerged as a new research field at the frontiers of nonequilibrium physics of complex systems, arising in superfluid helium, Bose-Einstein condensates, polaritonic condensates, and neutron stars \cite{BerloffSvistunov02,Finne2003,KobayashiPRL2005,Tsubota08book,SALMAN2009,Tsatsos16,Salman2016,Seo17,Wheeler_2021}, with applications ranging from atomtronics to astrophysics.

The turbulent regime in quantum fluids describes the complex, chaotic dynamics of interacting quantized vortices, which are topological defects. This definition is loose, and as we shall see, different criteria for its diagnosis have been put forward, including the scaling of the energy spectrum on the wave number, velocity autocorrelation functions, circulation statistics, etc. The situation is reminiscent of the use of various inequivalent definitions of quantum chaos, emphasizing complementary features brought out by a diversity of diagnostic tools (spectral statistics, semiclassical orbits, out-of-time order correlators, etc.) \cite{Nandy25}. 

QT was discovered in low-temperature physics studies of superfluid helium. However, the remarkable experimental progress in the mid-90s offered a new highly controllable and tunable platform for its exploration: Bose-Einstein condensates (BECs) of ultracold atomic gases \cite{Ueda10} and the closely related quantum fluids of light \cite{Carusotto13}. The experimental measurement of quantized vortices in BEC was crucial in probing their superfluid character. 
A BEC is characterized
by a macroscopic wave function that acts as a complex-valued order parameter 
and can be conveniently written in a polar decomposition $\Psi(\boldsymbol{r})=\sqrt{
\rho \left( \boldsymbol{r} \right) 
}\ e^{i\theta(\boldsymbol{r})}$ \cite{pitaevskii2003bose}. A defining feature of a superfluid is that the circulation $\Gamma[C]=\oint_C \boldsymbol{v}\cdot d\boldsymbol{\ell}$ (that is, the contour integral of the velocity around a given contour $C$) is quantized as a result of the single-valued character of the macroscopic wave function and the relation between the superfluid velocity $\boldsymbol{v}$ (equivalently, the vorticity $\boldsymbol{\omega}=\nabla\times\boldsymbol{v}$) and the condensate phase, $\boldsymbol{v}(\boldsymbol{r})=\hbar\nabla\theta(\boldsymbol{r})/m$, 
 
where $m$ is the mass of the particle.
QT differs from its classical counterpart in that quantum vortices are true topological defects and thus more stable than eddies, having conserved and quantized circulation \cite{Tsubota13,Tsatsos16}.

The phenomenology of turbulence varies significantly with the spatial dimension. 
In three spatial dimensions (3D), turbulence is characterized by a cascade of energy from large to small scales, with eddies breaking into ever-decreasing whirlpools. By contrast, two-dimensional (2D) turbulence is governed by the presence of an additional invariant of motion known as the enstrophy, and gives rise to an inverse energy cascade, with energy flowing from small to large scales. 
In the quantum case, the spatial dimensionality of the superfluid can be varied by tuning the anisotropy of the external confinement in BEC experiments. 
Notably, in 2D 
QT, 
enstrophy may not be conserved due to vortex-antivortex annihilation, and a direct energy cascade can therefore appear, as observed in numerical simulations \cite{Numasato2010,Chesler13} and 
BEC experiments \cite{Zhao25}.
In contrast, its conservation leads to the clustering of vortices in bigger patches, 
a behavior predicted by Onsager and experimentally reported in 
BECs 
\cite{Gau19,Johnstone19}
and exciton-polariton superfluids \cite{Panico23}.

\begin{figure*}[hptb]
    \centering
    \includegraphics[width=\linewidth]{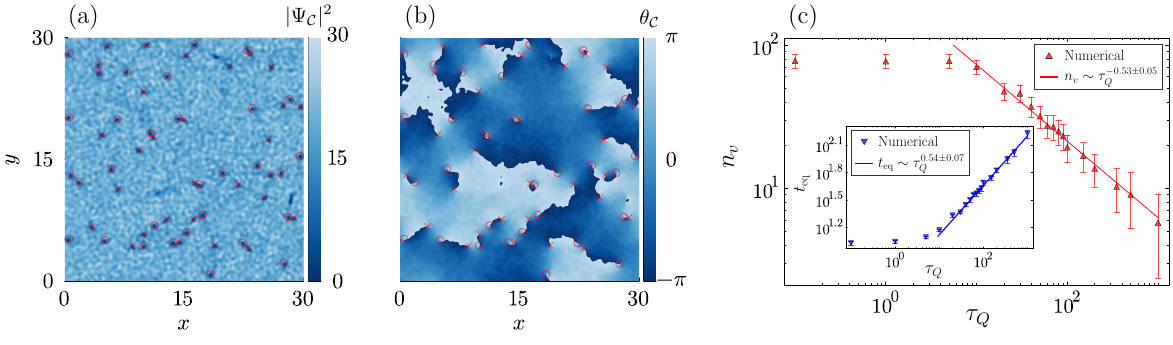}
    \caption{
    Typical condensate density and phase profiles at equilibration time, together with the vortex-number scaling. 
    Panel (a) shows the condensate density \(|\Psi_{\mathcal{C}}(\boldsymbol{r},t_{\textrm{eq}})|^{2}\) at equilibration time $t_{\textrm{eq}}$ following a quench of duration $\tau_{Q}=10$, while panel (b) displays the corresponding phase of $\Psi_{\mathcal{C}}$. Vortices with topological charge \(w=+1\) are marked by red crosses, and those with \(w=-1\) by circles; no vortices with \(\lvert w\rvert>1\) were observed. The main plot in panel (c) shows the
    vortex count 
    \( n_v \) 
    at \(t_{\textrm{eq}}\) versus quench time \(\tau_Q\), demonstrating the Kibble-Zurek power-law scaling for \(10\le\tau_Q\le1000\). The inset plots 
   \( {t}_{\textrm{eq}} \) 
    against \(\tau_Q\), where ${t}_{\textrm{eq}}$ is the equilibration time measured from the crossing of the critical point. Data are averaged over \(\mathcal{R}=1000\) realizations, with error bars indicating 1 standard deviation.
}
\label{fig:density_and_phase_psi_with_marked_vortices_and_KZ_scaling}
\end{figure*}
Several protocols are used to induce QT in a superfluid sample. They generally rely on continuous pumping of energy. A predominant approach involves sweeping a repulsive potential \cite{Kwon14,Reeves2022}. This procedure is often too violent, complicating the study of QT, as a result of the generation of a variety of excitations. 
As an alternative, one may wonder whether QT can emerge spontaneously, as topological defects emerge across a continuous phase transition. We refer to the QT that arises through finite-time spontaneous symmetry breaking as {\it spontaneous quantum turbulence} (SQT). In this context, the Kibble-Zurek mechanism (KZM) constitutes a powerful paradigm for describing universal critical dynamics \cite{Kibble76a,Kibble76b,Zurek96a,Zurek96b,DZ14}. It does so by exploiting equilibrium properties such as the divergence of the correlation length $\xi=\xi_0/|\varepsilon|^{\nu}$ and the relaxation time $\tau=\tau_0/|\varepsilon|^{z\nu}$ in the neighborhood of the critical point \cite{DZ14}. 
Here, $\nu$ and $z$ are the correlation-length and dynamic critical exponents, $\varepsilon = (\lambda - \lambda_c)/\lambda_c$ with $\lambda$ the control parameter (e.g., temperature) and $\lambda_c$ its critical value, while $\xi_0$ and $\tau_0$ are nonuniversal constants with dimensions of length and time, respectively.
For a linearized quench $\varepsilon=t/\tau_Q$, the KZM introduces the universal scaling of the nonequilibrium correlation length $\hat\xi$ with the quench time $\tau_Q$, i.e., $\hat\xi=\xi_0(\tau_Q/\tau_0)^{\nu/(1+z\nu)}$. For pointlike defects, such as solitons in a cigar-shaped BEC \cite{Damski_soliton_2010,Lamporesi13}, and vortices in a pancake BEC \cite{Weiler08,Chomaz15}, the KZM further predicts the formation of vortices at a density $\rho_{v}$ that scales universally with the quench time 
as $\rho_v  
\propto 1 / \hat{\xi}^2=\left( 1 / \xi_0^2 \right) \left( \tau_0 / \tau_Q \right)^{2 \nu/(1+z\nu)}$.
In addition, it dictates that the response of the system is delayed by the freeze-out timescale $\hat{t}=(\tau_0 \tau_Q^{z\nu})^{1/(1+z\nu)}$. In scenarios characterized by 
symmetry breaking that lead to vortex formation, the KZM has been widely tested in theoretical studies \cite{DRP11,Chesler15,Liu2020,Zeng23,Thudiyangal24,Wheeler2025,Kirkby2025}, with several experiments also supporting its role in superfluid formation and the BEC transition \cite{Ruutu1996,Weiler08,Navon15,Shin19,KimShin22}. 

In this Letter, we establish and describe SQT in a newborn BEC prepared by a thermal quench in finite time. We establish the Kolmogorov scaling of the incompressible kinetic energy 
spectrum and demonstrate its universal scaling for different values of the quench time when rescaled by the nonequilibrium correlation length predicted by the Kibble-Zurek mechanism. 

{\it Kibble-Zurek dynamics of the BEC transition---}
To describe the nonequilibrium condensate dynamics and the emergence of SQT, we consider the 
2D stochastic projected Gross-Pitaevskii equation \cite{Gardiner2003,blakie2008dynamics,Rooney2012,Rooney2014,Bradley2015,McDonald2020}. 
This formalism has been widely used to model the BEC transition, with numerous studies demonstrating its quantitative agreement with experimental observations \cite{Weiler08,Liu2020}.
Specifically, it describes the evolution of the low-energy coherent modes of a Bose gas via a complex order parameter $\Psi_{\mathcal{C}}$ which, 
in dimensionless units used throughout the manuscript, 
evolves according to 
\begin{equation}\label{2D_dimensionless_SPGPE_main}
    d{\Psi_{\mathcal{C}}} = \mathcal{P}_{\mathcal{C}} \left[ -(i + \gamma) \left( H_{\textrm{sp}} + g |\Psi_{\mathcal{C}}|^{2} -\mu\right) \Psi_{\mathcal{C}} \, dt + d \eta \right], 
\end{equation}
where $\mu$ is the chemical potential, $\gamma$ is the damping rate modeling energy dissipation due to the contact with the thermal bath, $g$ is the coupling strength of the delta-function interaction between bosons, and $d\eta$ is a complex Gaussian noise increment accounting for thermal fluctuations.
$H_{\textrm{sp}} = - \left( 1 / 2 \right) \nabla^{2} + V(\boldsymbol{r})$ denotes the single-particle Hamiltonian, with $V(\boldsymbol{r})$ the external trapping potential. In this work, we focus on a periodic homogeneous system and set $V(\boldsymbol{r})=0$. Further details of Eq.~(\ref{2D_dimensionless_SPGPE_main}) and its numerical implementation are provided in \cite{SM}. 

To drive the phase transition, we consider a linear quench of the chemical potential, $\mu(t) = \mu_i + \left( \mu_f - \mu_i \right)t/\tau_Q$, with $\mu_i$ and $\mu_f$ denoting its initial and final values, respectively. 
This protocol drives the system through the critical point $\mu_c$ at a finite rate $\tau_Q$, resulting in the formation of a BEC populated with vortices. A vortex is characterized by a quantized circulation of the superfluid velocity,
$
\oint \boldsymbol{v}(\boldsymbol{r},t)\cdot d \boldsymbol{\ell}=  
2 \pi
w
$, where $w\in\mathbb{Z}$ is the winding number.
The associated singularity of the velocity field enforces a vanishing condensate density at the vortex center, with the healing length determining the core size. 
An example of a single realization of the newborn BEC is shown in Fig.~\ref{fig:density_and_phase_psi_with_marked_vortices_and_KZ_scaling}, where vortices can be recognized by the depletion of the BEC density at their core. 
The associated phase map indicates that only the values $w=\pm1$ are generated in the course of the BEC transition. This is consistent with the fact that higher winding numbers are energetically costly and unstable against decay \cite{Tsubota08}.

Using the mean-field values of the critical exponents $\nu = 1/2$ and $z = 2$, the KZM predicts the scaling of the freeze-out time, which sets the timescale in which the growth of the order parameter lags behind the crossing of the critical point set by the quench of $\mu(t)$. The equilibration time, 
when the growth of the number of bosons in BEC changes from exponential to linear \cite{Chesler15,Thudiyangal24,SM}, 
is proportional to the freeze-out time and scales universally as 
$t_{\textrm{eq}} \propto \hat{t}\propto\tau_Q^{1/2}$. 
This scaling is confirmed by the numerical results shown in the inset of Fig.~\ref{fig:density_and_phase_psi_with_marked_vortices_and_KZ_scaling} (c), where 
$t_{\textrm{eq}} = ( 3.779 \pm 0.693 ) \tau_Q^{0.541 \pm 0.069}$. 
At the equilibration time, the KZM further predicts that the 
vortex 
number scales as $n_v \propto \tau_Q^{-1/2}$, in agreement with the fit shown in Fig.~\ref{fig:density_and_phase_psi_with_marked_vortices_and_KZ_scaling} (c), where $
n_v = \left( 249.036 \pm 44.999 \right) \tau_Q^{-0.534 \pm 0.047}
$.

Since we consider a 2D system, the role of the Berezinskii-Kosterlitz-Thouless (BKT) transition and the associated thermal vortices deserves discussion. Following the quasi-2D Bose gas experiment of Ref.~\cite{Chomaz15}, which demonstrated negligible BKT effects and verified KZM power laws, we work in the low-temperature limit (
of order 1 nK or less) to isolate the SQT arising from the symmetry-breaking phase transition.
In this regime, 
the expected number of thermal vortices is negligible \cite{Giorgetti2007}. This is further supported by the time evolution of the vortex count shown in the Supplemental Material \cite{SM}: $n_v$ decays monotonically to zero, indicating that all observed vortices are of KZM origin rather than thermally activated, consistent with the experimental findings of Ref.~\cite{Chomaz15}.

{\it Compressible and incompressible energy spectra---}
By introducing the density-weighted velocity field $\boldsymbol{u}(\boldsymbol{r})=|\Psi_{\mathcal{C}}(\boldsymbol{r})|\boldsymbol{v}(\boldsymbol{r})$ \cite{Nore1997PRL,Nore97,Ogawa2002,Kobayashi2005}, the kinetic energy of the newborn BEC takes the form 
$
E=
\left( 
1
/ 2 \right) 
\int |\boldsymbol{u}(\boldsymbol{r})|^2 \, d^2 \boldsymbol{r} . 
$
Using the canonical decomposition \cite{Nore1997PRL,Nore97,Ogawa2002,Kobayashi2005}, $\boldsymbol{u}$ can be separated into a compressible (irrotational) component $\boldsymbol{u}_c$, satisfying $\nabla \times \boldsymbol{u}_c = 0$, and an incompressible (solenoidal) component $\boldsymbol{u}_{i}$, for which $\nabla \cdot \boldsymbol{u}_{i} = 0$. The total kinetic energy is accordingly decomposed into its compressible and incompressible contributions,
\begin{equation}
E_{i,c} = \frac{
1
}{2} \int |\boldsymbol{u}_{i,c}(\boldsymbol{r})|^2 \, d^2 \boldsymbol{r}.
\label{eq:energy_cic}
\end{equation}
It is particularly useful to express Eq.~(\ref{eq:energy_cic}) in Fourier space, resulting in
$E_{i,c} = \int E_{i,c}(k) \, d k$, with
\begin{equation}
E_{i,c}(k) \coloneqq
\frac{k}{2}
\int 
|\boldsymbol{\tilde{u}}_{i, c} ( \boldsymbol{k} )|^{2} \, d \varphi_k,
\label{eq:energy_cic_fft}
\end{equation}
where $\boldsymbol{k} = k ( \cos \varphi_k, \sin \varphi_k )$ and $\boldsymbol{\tilde{u}}_{i,c}(\boldsymbol{k})$ is the Fourier transform of $\boldsymbol{u}_{i,c}(\boldsymbol{r})$ \cite{Bradley12}.

{\it Kolmogorov scaling of the energy spectrum in SQT---}
The occurrence of SQT in the newborn BEC is diagnosed by the Kolmogorov scaling of the incompressible kinetic energy spectrum 
$E_i \left( k \right)$ \cite{Tsubota13,Neely13}, which has been experimentally measured in 2D BECs \cite{Zhao25}. 
This law was originally derived for homogeneous isotropic turbulence where large eddies (with high Reynolds number) are continuously broken into smaller eddies (with small Reynolds number), and there exists a range of length scales (inertial range) where the details of energy injection and dissipation can be neglected and the eddies follow universal statistical properties \cite{Frisch1995,Barenghi2023textbook}. 
With the additional assumptions that, within this regime, 
the change of the kinetic energy per unit time 
is independent of $k$ and the viscosity does not affect $E_i(k)$, 
dimensional arguments imply a $k^{-5/3}$ scaling of the incompressible kinetic energy spectrum, independent of spatial dimensionality \cite{Vinen02,Barenghi2023textbook}. 
This universal law
is a direct reflection of the velocity-velocity autocorrelation function \cite{Kolmogorov41,Frisch1995}. 
At higher momenta, 
the 
inertial range
over which the Kolmogorov scaling extends is interrupted by a second power law, $E_i(k) \propto k^{-3}$, that reflects the short-range properties of the vortex core and is thus unrelated to turbulence \cite{Bradley12,Tanogami21}. 
As the radius of the vortex core is of the order of the condensate healing length $
\xi_h := \sqrt{ 
1 / (2
\mu_f)}
$ \cite{ginzburg1958,pitaevskii2003bose,Tanogami21}, the inertial range lies below $k < 2 \pi / \xi_h$.
At large scales, it is further bounded by the mean intervortex distance $l_{v}$ \cite{Bradley12}. The width of this range is therefore determined by the quench speed, which sets the energy-injection scale through the universal Kibble-Zurek (KZ) correlation length $\hat{\xi} \sim l_{v}$.
As a result, the extent of the Kolmogorov scaling shrinks for faster quenches and broadens for moderate ones. \begin{figure}[ptb]
\includegraphics[width=\columnwidth]{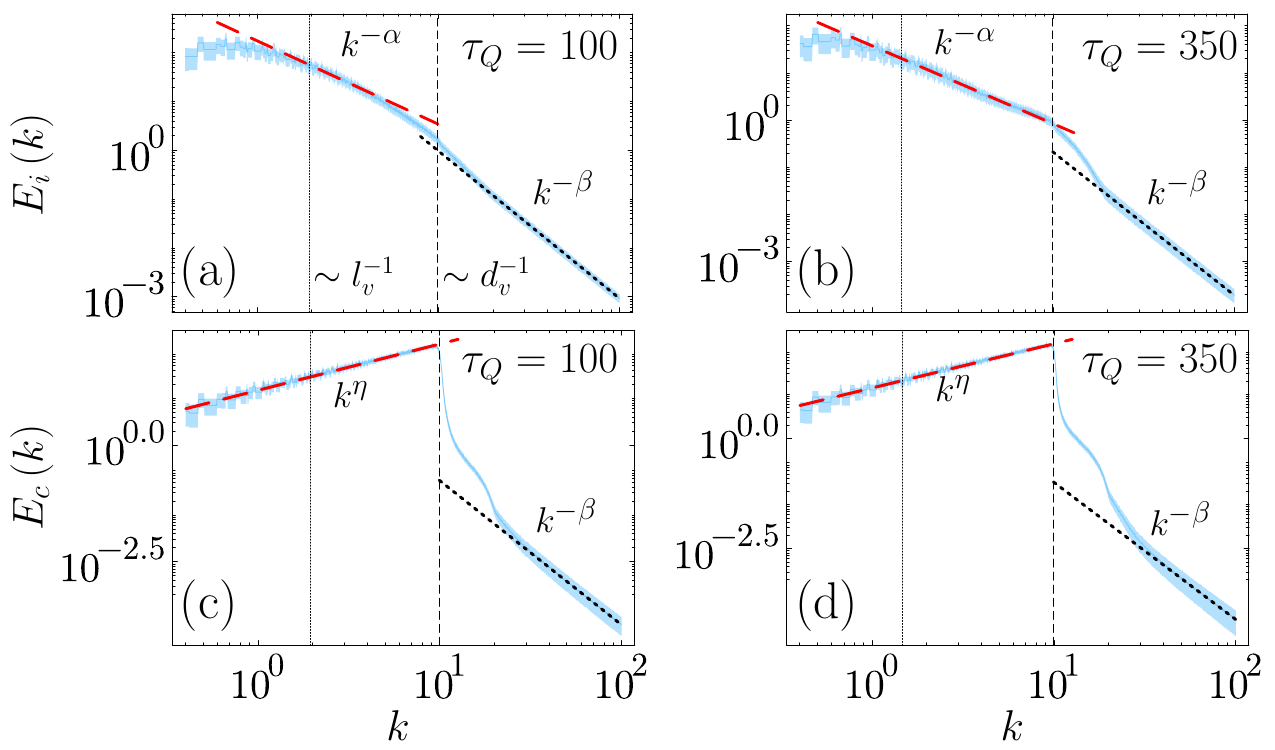}
\caption{
Incompressible and compressible kinetic energy spectra.
Panels (a) and (b) show the incompressible kinetic energy spectrum $E_{i}(k)$ at the equilibration time following quenches with 
$\tau_{Q}=100$ and $\tau_{Q}=350$, 
respectively, each averaged over $\mathcal{R}=1000$ independent noise realizations. The corresponding fitted power-law exponents are 
$\alpha = 1.710 \pm 0.079$, $\beta = 3.025 \pm 0.030$ for (a), and 
$\alpha = 1.654 \pm 0.031$, $\beta = 3.036 \pm 0.047$ 
for (b). Panels (c) and (d) display the compressible spectrum $E_{c}(k)$ for the same quench protocols, with fit parameters 
$(\eta, \beta)=(0.992 \pm 0.008, 3.090 \pm 0.110)$ and 
$(1.002 \pm 0.008, 3.111 \pm 0.223)$  
respectively.
The vertical dashed lines mark $k = 2\pi / d_v$, where $d_v = 4 \xi_h$ gives a good estimate of the vortex diameter in our system, 
while the vertical dotted lines indicate $k = 2 \pi / l_v$, with $l_v$ being the mean nearest intervortex distance. Shaded error bands correspond to 1 standard deviation.
}
\label{fig:E_vs_k}
\end{figure}
This effect is illustrated in Fig. \ref{fig:E_vs_k}, showing the incompressible energy spectra at equilibration for quenches of duration $\tau_{Q}=100$ in (a) and $\tau_{Q}=350$ in (b). 
The corresponding fits yield $E_i \left( k \right) = \left( 176.555 \pm 14.316 \right) k^{-1.710 \pm 0.079}$, $E_i \left( k \right) = \left( 1028.257 \pm 123.069 \right) k^{-3.025 \pm 0.030}$ in panel (a), and $E_i \left( k \right) = \left( 37.722 \pm 1.723 \right) k^{-1.654 \pm 0.031}$, $E_i \left( k \right) = \left( 226.968 \pm 42.383 \right) k^{-3.036 \pm 0.047}$ in (b), 
where in each panel the first and second expressions refer to the red dashed and black dotted lines, respectively. 
The power-law fitted exponents in the inertial range 
are thus in agreement with the Kolmogorov scaling value $5/3\simeq 1.67$ within the numerical uncertainty. 
In 
BECs the inertial range is typically narrow, spanning no more than one decade due to the lack of a wide separation between characteristic length scales \cite{Kobayashi2005, Tsubota13, Barenghi2023textbook}. 
This limitation can be mitigated by analyzing the velocity structure functions, which, due to their self-similar nature, allow the extension of the Kolmogorov scaling beyond the inertial range. 
In particular, in \cite{SM} we compute the longitudinal velocity increments $S_p(r)=\langle |[\boldsymbol{u}_i(\boldsymbol{R}+\boldsymbol{r})-\boldsymbol{u}_i(\boldsymbol{R})]\cdot
\boldsymbol{r} / r 
|^p \rangle$ and find clear power-law relations $S_p(r)\propto [S_3(r)]^{\zeta(p)}$, with exponents that deviate from the K41 prediction $\zeta(p)=p/3$ 
\cite{Kolmogorov41_first,Kolmogorov41,Benzi1993,Dubrulle1994,Ciliberto1994,Frisch1995,Krstulovic2016,Barenghi2023textbook,Zhao25}, and are more accurately described by the refined K62 model \cite{Kolmogorov_1962}.
While the KZM accurately describes the universal critical dynamics across the BEC transition, it does not account for other effects such as the annihilation of vortex-antivortex pairs, the losses of vortices at the edges of the trap, and coarsening. 
For slow quenches, the system evolves nearly adiabatically, resulting in suppressed vortex formation and diminished turbulence. As a consequence, the inertial range shrinks and the Kolmogorov scaling becomes less pronounced.

Although our main interest lies in the Kolmogorov scaling of the incompressible kinetic energy spectrum, we also plotted $E_c(k)$ at the equilibration time in Figs. \ref{fig:E_vs_k}(c) and \ref{fig:E_vs_k}(d). 
For a 2D system at thermodynamic equilibrium, it is reported that $E_c(k) \propto k$ for small $k$ \cite{Numasato2010}. 
In addition, since $E(k)$ follows the $k^{-3}$ law at large $k$ \cite{Nore97,Krstulovic10}, $E_c(k)$ is expected to follow the same power law. Our numerical results in Figs. \ref{fig:E_vs_k}(c) and \ref{fig:E_vs_k}(d) are in agreement with these expectations. 
\begin{figure}[ptb]
\includegraphics[width=0.8\columnwidth]{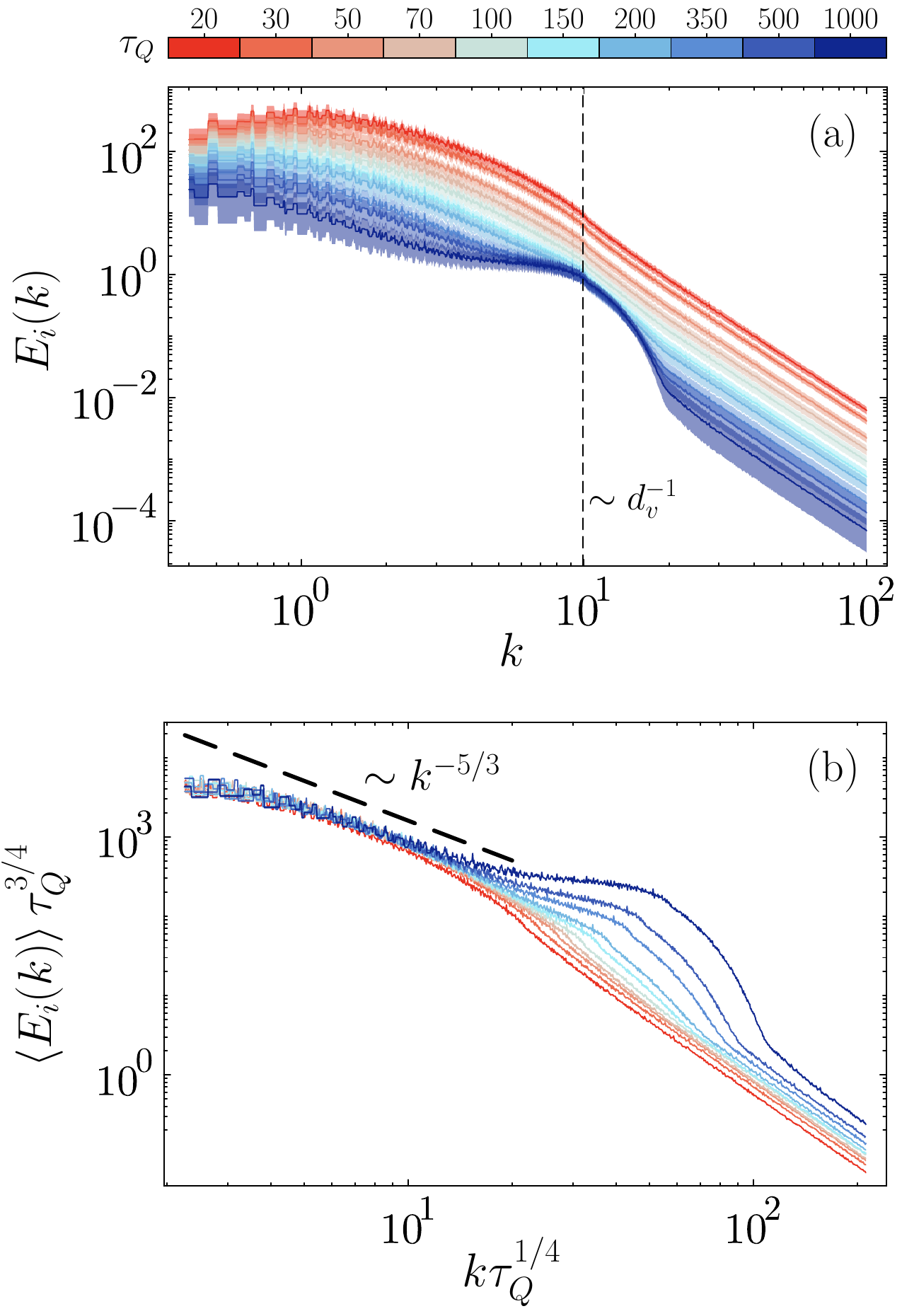}
\caption{
Kibble-Zurek universality of the incompressible kinetic energy spectrum.
Panel (a) shows $E_i \left( k \right)$ at equilibration time for various $\tau_{Q}$ values, each averaged over $\mathcal{R}=1000$ stochastic realizations. 
The 
dashed 
vertical line marks 
$k = 2\pi / d_v$ with the vortex diameter estimated to be $d_v = 4 \xi_h$. 
Panel (b) displays 
$\left\langle E_i \left( k \right) \right\rangle \tau_Q^{
3/4
}
$  
as a function of the scaled momentum $k \tau_Q^{1/4} \propto k \hat{\xi}$. 
Shaded error bands indicate 1 standard deviation in (a) and 95\% confidence interval in (b). 
}
\label{fig:Ei_vs_k_KZM}
\end{figure}

{\it Kibble-Zurek universality of SQT---}
Having established that the Kolmogorov scaling describes thermal quenches for two different values of $\tau_Q$, we now demonstrate the nonequilibrium universality dictated by KZM in SQT.
To this end, let us consider the total incompressible kinetic energy 
$E_i$. 
Since the density-weighted velocity field $\boldsymbol{u}_{i}=\sqrt{\rho}\,\boldsymbol{v}_{i}$ is concentrated around the vortex cores, $E_i$ can be approximated as a sum of isolated vortex contributions,
$
E_i 
\simeq
\left( 
1
/ 2 \right) 
n_v 
\!\int_{\mathcal{D}_{\hat{\xi}}}\!
\left\vert 
\boldsymbol{u}_v(\boldsymbol{r})
\right\vert^2 
\,d^2\boldsymbol{r},
$
where \(
\left\vert 
\boldsymbol{u}_v(\boldsymbol{r})
\right\vert 
=\sqrt{\rho(\boldsymbol{r})}\, 
/ r
\), and the integration domain \(\mathcal{D}_{\hat{\xi}}\) is set by the KZ correlation length \(\hat{\xi}\).  
At equilibration, the background density \( \rho \simeq \mu(t_{\textrm{eq}}) / g \propto 
\tau_Q^{- 1 / \left( 1 + z \nu \right)}
\) and \( \hat{\xi} \propto 
\tau_Q^{\nu / \left( 1 + z \nu \right)}
\), so each vortex contributes \( \int_{\mathcal{D}_{\hat{\xi}}} 
\left\vert 
\boldsymbol{u}_v 
\right\vert^2 
\, d^2\boldsymbol{r} \propto 
\tau_Q^{- 1 / \left( 1 + z \nu \right)}
\) up to logarithmic corrections \cite{SM}. Combining this with the KZ vortex number scaling \( n_v \propto 
\tau_Q^{- 2 \nu / \left( 1 + z \nu \right)}
\) yields \( E_i \propto 
\tau_Q^{- \left( 1 + 2 \nu \right) / \left( 1 + z \nu \right)}
\) as confirmed numerically in 
the End Matter. 
Next, we consider the spectral density $E_i(k)$. Since the Kibble-Zurek length $\hat{\xi}$ sets the characteristic momentum scale at equilibration, it is natural to assume the scaling ansatz $E_i(k,\tau_Q) = A(\tau_Q) F(k\hat{\xi})$, where $F$ encodes the spectral shape and $A(\tau_{Q})$ is a $k$-independent amplitude. Using $E_i \propto 
\tau_Q^{- \left( 1 + 2 \nu \right) / \left( 1 + z \nu \right)}$ together with $\hat{\xi} \propto \tau_Q^{\nu / \left( 1 + z \nu \right)}$ gives $A(\tau_Q) \propto \tau_Q^{- \left( 1 + \nu \right) / \left( 1 + z \nu \right)}$, implying a collapse of the spectra across different quench rates under the rescaling
\begin{equation}
k \to k\, \tau_Q^{\nu / \left( 1 + z \nu \right)}
, \qquad 
E_i(k) \to E_i(k)\, \tau_Q^{\left( 1 + \nu \right) / \left( 1 + z \nu \right)}
,
\label{Ei_scaling}
\end{equation}
as shown in Fig.~\ref{fig:Ei_vs_k_KZM} for the mean-field values of the critical exponents. 
A universal collapse also emerges for the compressible spectrum $E_c(k)$, under a distinct rescaling, at low temperature where phonon emission from vortex-antivortex annihilation 
\cite{Kumar2025} 
dominates over thermal density fluctuations (see \cite{SM} for details).
In addition, for $k$ exceeding $d_v^{-1}$, the lack of collapse for $E_i(k)$ when rescaled by the KZ variables in the $k^{-3}$ scaling regime is expected as it is governed by an equilibrium quantity, the vortex core radius, that is independent of the quench rate. 

Deviations from the universal collapse can also be expected for fast quenches (e.g., $\tau_Q \leq 10$), for which the KZM breaks down \cite{Zeng23,Xia24} due to the saturation of the defect density shown in Fig. \ref{fig:density_and_phase_psi_with_marked_vortices_and_KZ_scaling} (c). However, such deviations may be taken into account by using the associated correlation length and energy scale, which scale universally with the quench depth rather than the quench time.
Naturally, other nonuniversal processes beyond KZM and its extensions, governing the evolution after the formation of the BEC, such as atom losses, vortex-antivortex annihilation, and coarsening are expected to suppress the Kolmogorov scaling and SQT. 

{\it Discussion and conclusion---}
In summary, we have shown that superfluid formation triggered by a finite-time quench gives rise to SQT. To this end, we have established that the incompressible kinetic energy of the emergent nonequilibrium BEC exhibits Kolmogorov scaling with respect to the wave vector. Moreover, using the nonequilibrium correlation length predicted by KZM and the associated energy scale, we have shown that the Kolmogorov scaling admits a universal collapse  
for different values of the quench time.
Our findings highlight the rich interplay between distinct yet complementary notions of universality: the universal scaling of energy spectra in turbulent flows and the universal defect dynamics across continuous phase transitions.

{\it Note added---}
During the completion of this work, Ref. \cite{Yang2025} reported the study of a 4D superfluid with Kolmogorov scaling for the total energy, rather than the incompressible energy in the 2D case we discuss.

{\it Acknowledgments---}
M. M. is grateful to Andr\'{a}s Grabarits, Simon Fisher, 
Martin Gazo, and Zoran Hadzibabic for valuable discussions. S. S. would like to further thank Sol Kim and Yong-il Shin for insightful comments, and the Institut d'\'Etudes Scientifiques de Carg\`ese, for support and hospitality during the program Bridges over turbulent matters - Navigating across observations, concepts and models. The authors acknowledge financial support from the Luxembourg National Research Fund under Grant No. C22/MS/17132060/BeyondKZM. M. T. would like to thank Anusandhan National Research Foundation (ANRF), Government of India, for the financial support through the Prime Minister Early Career Research Grant with Grant No. ANRF/ECRG/2024/003150/PMS, and Christ University for funding the research through the seed grant, Sanction No. CU-ORS-SM-24/94. The numerical simulations presented in this work were carried out using the HPC resources of the University of Luxembourg (ULHPC). 

{\it Data Availability---}
The data that support the findings of this article are openly available \cite{Shinn_2025_Zenodo_SQT}. 


\bibliographystyle{apsrev4}
\let\itshape\upshape
\normalem
\bibliography{defects_QT}

\newpage

\appendix
\onecolumngrid

\section{End Matter}

\twocolumngrid
{\it Universality of the incompressible kinetic energy---}
\begin{figure}[pb]
\includegraphics[width=\columnwidth]{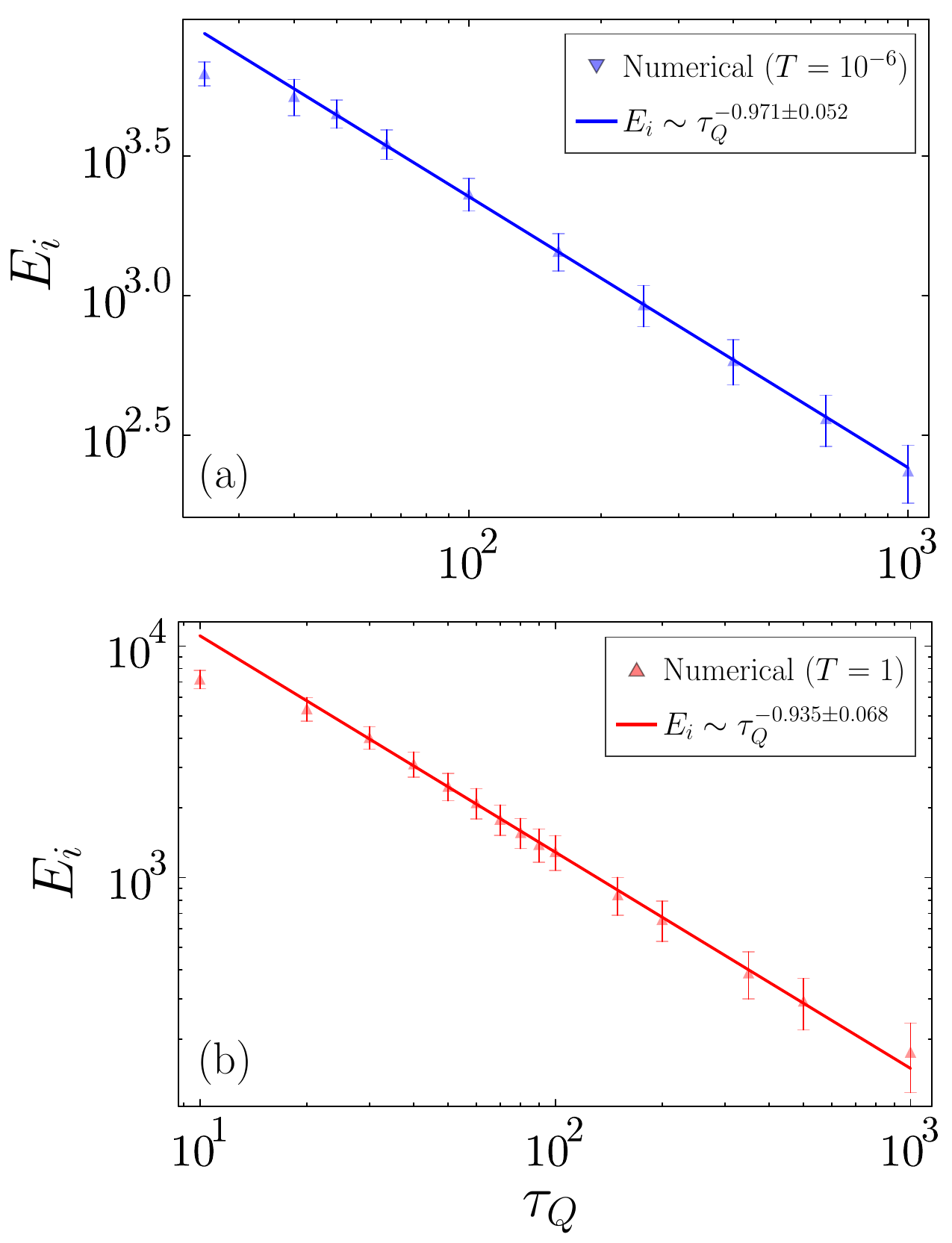}
\caption{
Kibble-Zurek universality of the incompressible kinetic energy. 
Panels (a) and (b) show the incompressible kinetic energy $E_i$ at the equilibration time for various quench times $\tau_Q$ at temperature $T = 10^{-6}$ and $T = 1$, respectively. Each data point is averaged over (a) $\mathcal{R}=2000$ and (b) $\mathcal{R}=1000$ stochastic realizations. In both panels, error bars denote 1 standard deviation. 
}
\label{fig:Ei_integ_Tq_KZM}
\end{figure}
In the main text, we argued that the incompressible kinetic energy at equilibration 
time 
$
E_i 
\coloneqq 
\left( 
1
/ 2 \right) 
\int 
\left\vert 
\boldsymbol{u}_i \left( \boldsymbol{r}, t_{\textrm{eq}} \right) 
\right\vert^2 
\; 
d^2 \boldsymbol{r} 
= 
\int_{0}^{\infty} 
E_i \left( k \right) 
\; 
d k 
$, scales with the quench duration as
\begin{equation}\label{eq:scaling_E_i_nu_z}
E_i \propto 
\tau_Q^{- \left( 1 + 2 \nu \right) / \left( 1 + z \nu \right)}.
\end{equation} 
This follows from the observation that $E_i$ can be approximated by a sum over the $n_v$ vortex contributions, 
$
E_i 
\simeq 
(
1
/2)\, n_v \!\int_{\mathcal{D}_{\hat{\xi}}}\! |\boldsymbol{u}_v(\boldsymbol{r})|^2\, d^2\boldsymbol{r}
$, where the integration domain $\mathcal{D}_{\hat{\xi}}$ is set by the KZ length $\hat{\xi}$, and the weighted velocity field of a single vortex is $
\boldsymbol{u}_v(\boldsymbol{r}) = \sqrt{\rho(\boldsymbol{r})}\,
\boldsymbol{e}_{\varphi}
\, / r
$ 
with $\boldsymbol{e}_{\varphi}$ being the unit vector along the azimuthal angle. 
We evaluate the single-vortex integral by assuming a vanishing condensate density within the core of radius $r_{v}$, and a uniform background value $\rho=\mu/g$ outside. This yields 
\begin{equation}
    \int_{\mathcal{D}_{\hat{\xi}}}\!\left\vert \boldsymbol{u}_v(\boldsymbol{r})\right\vert^2 \,d^2\boldsymbol{r}
    \simeq
    2 \pi
    \rho 
    \int_{r_{v}}^{\hat{\xi}}\frac{1}{r}\, dr= 
    2 \pi
    \rho
    \ln\left(\frac{\hat{\xi}}{r_v}\right).  
\end{equation}
The factor $\ln(\hat{\xi}/r_v)$ grows only logarithmically with $\hat{\xi}$, making it a subleading correction to the power law (\ref{eq:scaling_E_i_nu_z}), which is governed by the scaling of $n_v$ and $\rho$ at equilibration time, as discussed in the main text. 
We tested the prediction in Eq. (\ref{eq:scaling_E_i_nu_z}) numerically for two different temperature values in Fig. \ref{fig:Ei_integ_Tq_KZM}, and the results support this scaling within numerical uncertainty. The fitted power laws in Fig. \ref{fig:Ei_integ_Tq_KZM} are  $E_i = (19.774 \pm 5.010) \times 10^4 \,\tau_Q^{-0.971 \pm 0.052}$ for panel (a) and $E_i = (9.543 \pm 2.951) \times 10^4 \,\tau_Q^{-0.935 \pm 0.068}$ in panel (b), confirming the predicted $\tau_Q^{-1}$ behavior 
based on the mean-field values of the critical exponents $\nu = 1/2$ and $z = 2$. 
For all fits with exact independent variables, we use weighted least squares with weights given by the inverse variances.

\newpage
\clearpage

\title{---Supplementary Material--- \\Spontaneous Quantum Turbulence in a Newborn Bose-Einstein Condensate via the Kibble-Zurek Mechanism
}
\maketitle
\onecolumngrid

\section{Two-dimensional stochastic projected Gross-Pitaevskii equation}
The stochastic projected Gross-Pitaevskii equation (SPGPE) is a description based on $c$-field methods \cite{blakie2008dynamics}, consisting of dividing the system's modes into a coherent region ($\mathcal{C}$), which includes the low-lying, highly populated levels, and an incoherent region of high-energy, sparsely occupied states that act as a thermal reservoir. The evolution of $\mathcal{C}$ is captured by a classical field $\Psi_{\mathcal{C}}$, which evolves according to
\begin{equation}\label{2D_dimensionless_SPGPE}
    d{\Psi_\mathcal{C}} = \mathcal{P}_{\mathcal{C}} \left[ -(i + \gamma) \left( H_{\textrm{sp}} + g |\Psi_\mathcal{C}|^{2} -\mu(t) \right) \Psi_\mathcal{C} \, dt + d \eta \right], 
\end{equation}
where length, time, and temperature are measured in units of $l \coloneqq \sqrt{\hbar^2 / \left( m E_{\textrm{sc}} \right)}$, $\hbar / E_{\textrm{sc}}$, and $E_{\textrm{sc}} / k_B$, respectively. Here $E_{\textrm{sc}}$ sets the energy scale, $m$ is the mass of a boson, $\hbar$ the reduced Planck constant, $k_B$ the Boltzmann constant, $\mu$ the chemical potential, $\gamma$ the dissipation rate, 
$
g 
= 
2 \sqrt{2 \pi} a_s / l_{\perp} 
$ 
the coupling strength of the repulsive delta-function interaction between bosons 
in a quasi-two-dimensional system \cite{pitaevskii2003bose,Bradley2015}, with 
$a_s$ the s-wave scattering length and 
$l_{\perp}=\sqrt{\hbar/(mw_{\perp})}$ the harmonic-oscillator length associated with the transverse confinement. 
From now on, we will use our scaled units throughout the supplementary material. 
$H_{\textrm{sp}}=-(1/2)\nabla^{2}+V(\boldsymbol{r})$ denotes the single-particle Hamiltonian, and its eigenfunctions $\phi_{n}(\boldsymbol{r})$ serve as a convenient basis to describe the $\mathcal{C}$ region. 
The noise term $d\eta$ is a complex Wiener process, with an autocorrelation function proportional to the temperature $T$. Specifically, it satisfies
\begin{equation}
\langle d \eta (\boldsymbol{r}, t) \rangle = 0, \quad \langle d \eta (\boldsymbol{r}, t) d \eta^{*}(\boldsymbol{r}', t) \rangle = 2T\gamma\, \delta_{\mathcal{C}}(\boldsymbol{r} - \boldsymbol{r}')\, dt,
\end{equation}
where $\delta_{\mathcal{C}}(\boldsymbol{r}-\boldsymbol{r}^{\prime})=\sum_{n\in \mathcal{C}}\phi_{n}(\boldsymbol{r})\phi^{*}_{n}(\boldsymbol{r}^{\prime})$
is the delta function restricted to the low-energy subspace. This subspace is spanned by the set of eigenfunctions $\{\phi_{n}\}_{n \in \mathcal{C}}$ with energies $\epsilon_{n}$ below a properly chosen cut-off $\epsilon_{cut}$. 
Finally, $\mathcal{P}_{\mathcal{C}}$ denotes the projector operator,
which restricts the dynamics to the low-energy subspace.
Specifically,
$\mathcal{P}_{\mathcal{C}}$ acts on a generic field $f(\boldsymbol{r})$ by projecting it onto the $\mathcal{C}$ region \cite{blakie2008dynamics}:
\begin{equation}\label{action_Projector_C}
    \mathcal{P}_\mathcal{C} f(\boldsymbol{r}) = \sum_{n:\epsilon_{n}<\epsilon_{\text{cut}}} \phi_{n}(\boldsymbol{r}) \int f(\boldsymbol{r}^{\prime}) \phi^{*}_{n}(\boldsymbol{r}^{\prime}) \, d^{2}\boldsymbol{r}^{\prime}.
\end{equation} 
The action of the projector is thus naturally implemented in the eigenbasis of the single-particle Hamiltonian $H_{\textrm{sp}}$, by setting a maximum eigenmode and ensuring that the dynamics does not develop components above this cutoff. The field $\Psi_{\mathcal{C}}$ is then conveniently represented as a sum over the corresponding basis states
\begin{equation}\label{decompose_Psi_C_eq}
    \Psi_{\mathcal{C}}(\boldsymbol{r},t)=\sum_{n\in \mathcal{C}} c_{n}(t)\phi_{n}(\boldsymbol{r}),
\end{equation}
with the weights $c_{n}$ encoding the particle occupation number in the $n$-th mode. Substituting this expansion into (\ref{2D_dimensionless_SPGPE}) yields the evolution equations for the coefficients $c_{n}$.
To integrate these equations numerically, we rely on the Julia library \texttt{DifferentialEquations.jl} \cite{rackauckas2017differentialequations}, employing a stability-optimized stochastic Runge-Kutta algorithm (SOSRA) for time propagation \cite{rackauckas2017adaptive,rackauckas2020stability}.
The simulations are performed in a periodic $L \times L$ homogeneous box (i.e., $V(\boldsymbol{r})=0$), with $\Psi_{\mathcal{C}}=0$ as the initial condition. The system is then relaxed for $t_{r}$ time units before beginning the quench.
Unless otherwise specified, we use the parameter values in Table \ref{tab:tab_param_vals}, motivated by the choices in \cite{Thudiyangal24}.

\begin{table}[h]
\centering
\setlength{\tabcolsep}{6pt}
\renewcommand{\arraystretch}{1.3} 
\caption{Simulation parameters }
\label{tab:tab_param_vals}
\begin{tabular}{cccccccc}
\hline\hline
$L$ & $t_r$ & $\mu_i$ & $\mu_f$ & $g$ & $\gamma$ & $T$ & $\epsilon_{\mathrm{cut}}$ \\
\hline
30  & 10    & 0.1     & 20      & 1   & 0.03     & 1   & $2.5\,\mu_f$ \\
\hline\hline
\end{tabular}
\end{table}

\subsection{Determination of the equilibration time}
\begin{figure}[hptb]
\centering
\includegraphics[width=0.7 \linewidth]{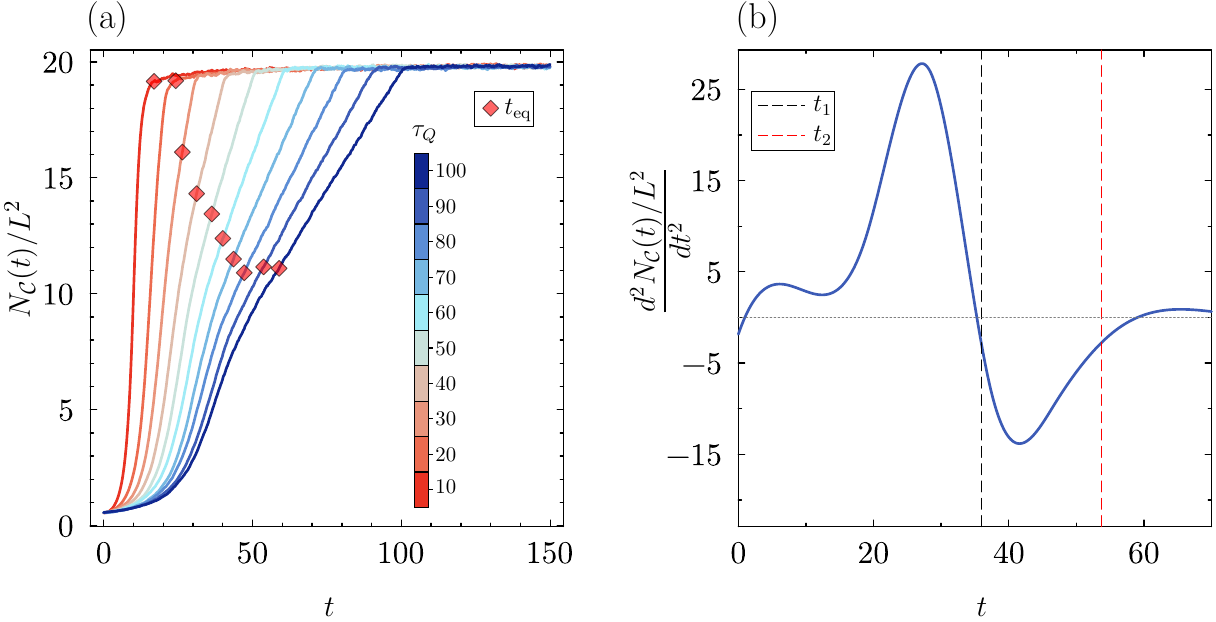}
\caption{Determination of the equilibration time. Panel (a) shows the time evolution of the $\mathcal{C}$-region number density (i.e., $N_{\mathcal{C}}(t)/L^2 = (1 / L^2) \int |\Psi_{\mathcal{C}}(\boldsymbol{r},t)|^2 \, d^{2}\boldsymbol{r}$) for quench times $\tau_{Q}$ from $10$ to $100$. The equilibration time $t_{\textrm{eq}}$ is determined by considering the second derivative of $N_{\mathcal{C}}/L^2$ and identifying the time interval $t_{1}\leq t\leq t_{2}$ such that $d^2 (N_{\mathcal{C}}/L^2) \left( t \right) / d t^2 \le - \Delta$ for $t_1 \le t \le t_2$. In particular, we set $\Delta$ to be $10\%$ of the maximum value of $d^2 (N_{\mathcal{C}}/L^2) \left( t \right) / d t^2$. This is shown for the case $\tau_{Q}=90$ in panel (b).
Finally, the equilibration time is defined as $t_{\textrm{eq}}=t_2$.
}
\label{fig:NC_sec_NC_teq}
\end{figure}

The equilibration time $t_{\textrm{eq}}$ is identified by analyzing the time evolution of the norm of the order parameter, $N_{\mathcal{C}}/L^{2} =(1/L^{2})
\int |\Psi_{\mathcal{C}}(\boldsymbol{r})|^{2} \, d^{2}\boldsymbol{r}$.
In the spirit of \cite{Chesler15}, we locate the global maximum of $d^{2}(N_{\mathcal{C}}/L^2)/dt^{2}$ and identify the subsequent interval $(t_1, t_2)$ where $d^{2}(N_{\mathcal{C}}/L^2)/dt^{2} \le -\Delta$, with $\Delta$ set to $10\%$ of the maximum. We then take $t_{\textrm{eq}}=t_{2}$. The procedure is illustrated in Fig. \ref{fig:NC_sec_NC_teq}.

The equilibration time measured from the critical crossing scales with the Kibble-Zurek (KZ) freeze-out time as
\begin{equation}
\bar{t}_{\mathrm{eq}} \coloneqq t_{\mathrm{eq}} - t_c \propto \hat{t} \propto \tau_Q^{z \nu /(1 + z \nu)},
\end{equation}
where \(t_c = \tau_Q (\mu_c - \mu_i) / (\mu_f - \mu_i)\) denotes the time at which the system crosses the critical point, and \(\mu_c \coloneqq \mu(t_c)\) is the corresponding value of the chemical potential.
This universal scaling is confirmed in the mean-field regime by the numerical results presented in the main text, which yield
\begin{equation}
t_{\mathrm{eq}} = (3.779 \pm 0.693)\,\tau_Q^{0.541 \pm 0.069} + (0.078 \pm 0.059)\,\tau_Q.
\end{equation}

\subsection{Condensate fraction}
\begin{figure}[hptb]
\centering
\includegraphics[width=0.5\linewidth]{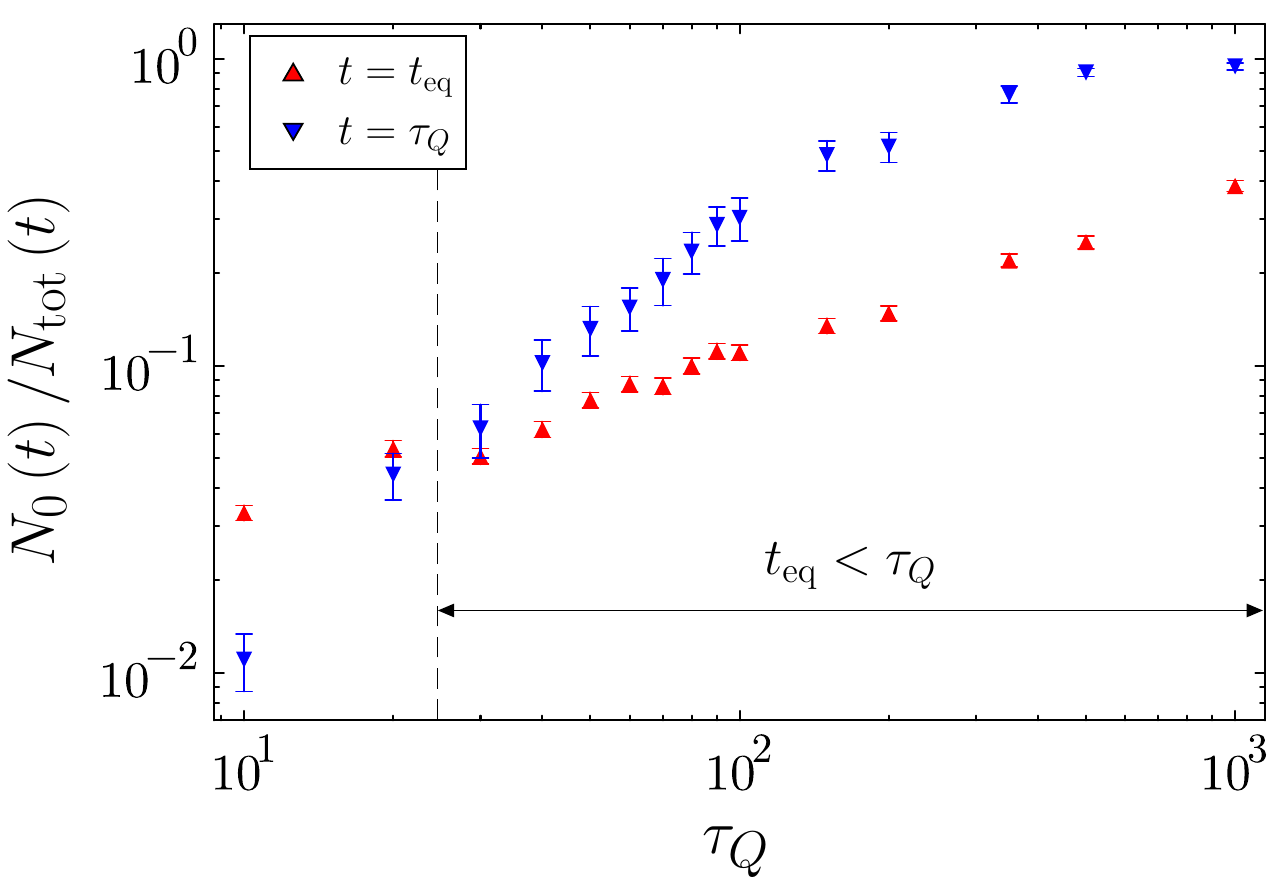}
\caption{Condensate fraction at $t = t_{\textrm{eq}}$ and $t = \tau_Q$. 
The condensate fraction is shown at equilibration time $t_{\textrm{eq}}$ and at the end of the quench ($t = \tau_Q$). 
The values at $t_{\textrm{eq}}$ are averaged over $\mathcal{R}=1000$ stochastic realizations (same dataset as in the main text), while those at $t = \tau_Q$ are averaged over $\mathcal{R}=100$ realizations. Error bars denote 95\% confidence interval. 
}
\label{fig:c_frac_at_teq_Tq}
\end{figure}

Within the SPGPE description, the condensate fraction (defined as the ratio between the number of bosons in the 
Bose-Einstein condensate (BEC) 
state and the total number of bosons in the system) can be computed efficiently by means of the Penrose-Onsager criterion \cite{Penrose1956}. In particular, the occupation $N_{0}$ of the condensate mode is given by the largest eigenvalue of the single-particle density matrix
\begin{equation}
    \rho(\boldsymbol{r},\boldsymbol{r}^{\prime}) = \langle \Psi_{\mathcal{C}}(\boldsymbol{r}) \Psi_{\mathcal{C}}^{*}(\boldsymbol{r}^{\prime}) \rangle,
\end{equation}
where $\langle \cdot \rangle$ denotes the ensemble average. In practice, this quantity is most conveniently evaluated in the eigenmode basis of the single-particle Hamiltonian \cite{blakie2008dynamics}.
The condensate fraction is then computed as $N_{0}/N_{\text{tot}}$, where $N_{\text{tot}}$ denotes
the total number of bosons in the system. Specifically, $N_{\text{tot}}=N_{\mathcal{C}} + N_{\mathcal{I}}$, with $N_{\mathcal{C}}=\int |\Psi_{\mathcal{C}}(\boldsymbol{r})|^2\, d^2{\boldsymbol{r}}$ the population of the coherent ($\mathcal{C}$) low-energy sector, and $N_{\mathcal{I}}$ the number of thermal bosons belonging to the incoherent ($\mathcal{I}$) high-energy region. The atoms in $\mathcal{I}$ are assumed to be non-interacting and thermalized, and their number can therefore be determined using the Bose-Einstein distribution~\cite{Liu2020}.
Fig. \ref{fig:c_frac_at_teq_Tq} shows that the condensate fraction increases as the quench speed decreases, consistent with slower quenches generating fewer excitations and resulting in a more coherent condensate at the end of the quench (for reference, $N_0 \left( \tau_Q \right) / N_{\textrm{tot}} \left( \tau_Q \right) = 0.945
\pm 0.024
$ for $\tau_Q = 1000$, averaged over 100 stochastic realizations). We also note that for quench times $\tau_Q$ below a threshold lying between 20 and 30, the quench is rapid enough that the equilibration time exceeds the quench duration. Consequently, for these $\tau_Q$ values the condensate fraction is larger at $t_{\textrm{eq}}$ than at $t = \tau_Q$.

\subsection{Vortex detection algorithm}
To determine the positions of the vortex cores at equilibration time, we locate the zeros of the condensate density $\rho_{\mathcal{C}}(\boldsymbol{r},t_{\textrm{eq}})=|\Psi_{\mathcal{C}}(\boldsymbol{r},t_{\textrm{eq}})|^2$ and check the associated winding number.
Specifically, we discretize the $L\times L$ system onto a $N_g \times N_g$ grid, with $N_g = 3000$, and identify each lattice site where
$\rho_\mathcal{C} \left( \boldsymbol{r}, t_{\textrm{eq}} \right)$ has a local minimum with value below $0.1\mu_f / g$. Starting from these initial guesses, we refine the locations of the zeros of $\rho_\mathcal{C}$ by iteratively applying the Newton-Raphson algorithm, until the condition $\rho_\mathcal{C} \left( \boldsymbol{r}\right) \le 10^{-5}$ is fulfilled \cite{Villois_2016}.
For each vortex candidate position, we calculate the winding number by integrating the superfluid velocity along a circle of radius \(r_{\text{test}} = 0.06\,\xi_h\) centered on it, confirming whether it is a true vortex core.

\section{Velocity structure functions analysis and extended self-similarity}
In the main text, we have shown that a BEC created via a thermal quench develops a turbulent superfluid state characterized by the celebrated $k^{-5/3}$ Kolmogorov scaling in the incompressible kinetic energy spectrum. 
This power law holds within the inertial range (IR), corresponding to wavenumbers associated with length scales between the typical nearest inter-vortex separation (which sets the energy injection scale controlled by the quench rate) and the vortex core size, beyond which the spectrum reflects the static vortex structure rather than turbulence. 
\begin{figure}[hpb]
\centering
\includegraphics[width=0.8 \linewidth]{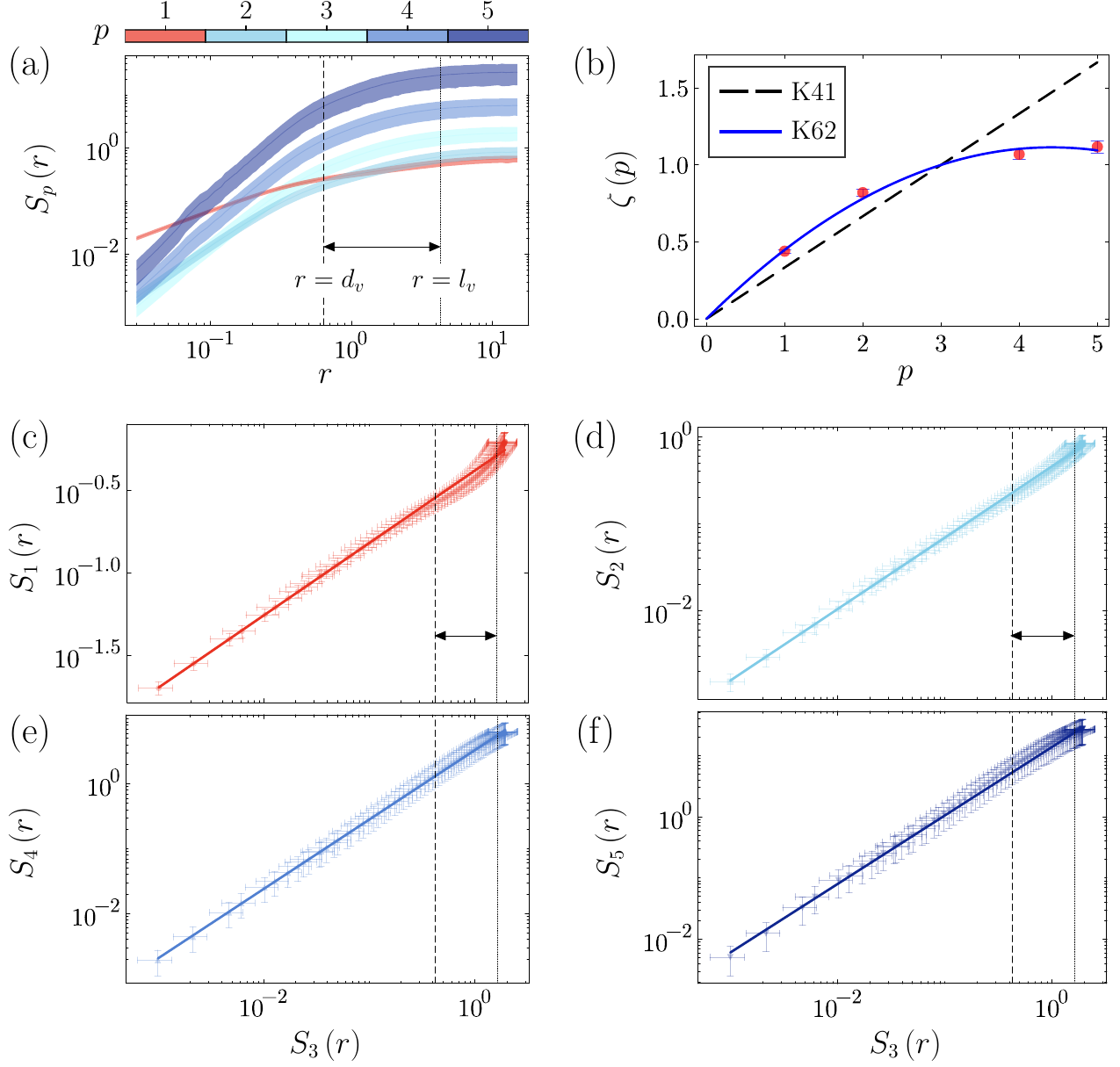}
\caption{Velocity structure functions and extended self-similarity. 
(a) Velocity structure functions $S_p (r)$ at equilibration time for a quench of duration  $\tau_Q = 350$ at temperature $T = 1$, averaged over $\mathcal{R}=1000$ independent noise realizations. Here, the vertical lines mark the mean nearest inter-vortex distance $l_{v}$, and the vortex diameter $d_v=4 \xi_h$, with $\xi_h$ being the healing length.
(b) Scaling exponents $\zeta(p)$ obtained from the relation $S_p(r) \propto \left\lbrack S_3(r) \right\rbrack^{\zeta(p)}$.
(c - f) $S_p(r)$ as a function of $S_3(r)$, where the vertical lines delimit the region $S_3(d_v) \le S_3(r) \le S_3(l_v)$. Solid lines are fits with $S_p(r) \propto \left\lbrack S_3(r) \right\rbrack^{\zeta(p)}$ 
using the whole range of $r$ in (a), i.e., $0.03 \le r \le 15$, 
where the fitted values of $\zeta(p)$ are the ones shown in (b). 
In all panels except (b), error bands and error bars represent one standard deviation. 
In panel (b), $\zeta(p)$ is computed by accounting for uncertainties in both $S_p(r)$ and $S_3(r)$, and the resulting error bars indicate the standard errors (i.e., the square roots of the diagonal elements of the covariance matrix) obtained from the orthogonal distance regression.
}
\label{fig:ESS_Tq_350}
\end{figure}
The extent of this range usually spans about one decade in $k$ or less, depending on the quench rate, reflecting the limited separation of characteristic length scales in atomic BECs.
In particular, the ratio of the system size $L$ to the healing length $\xi_h$ is typically $L/\xi_h \sim O(70)$ in experiments \cite{Chomaz15}, while in our simulation it is $L/\xi_h = 30\sqrt{2\mu_f} = 189.737$.
To mitigate this limitation and obtain a broader scaling regime, we resort to the principle of extended self-similarity of the velocity structure functions.
In particular, we consider the $p$-th order longitudinal velocity structure function $S_p \left( r \right)$, defined as 
\begin{equation}
S_{p}(r) = \Big\langle \big\vert[\boldsymbol{u}_{i}(\boldsymbol{R}+\boldsymbol{r}) - \boldsymbol{u}_{i}(\boldsymbol{R})]\cdot 
\boldsymbol{e}_r 
\big\vert^{p} \Big\rangle,
\end{equation}
where $\boldsymbol{u}_{i}$ denotes the incompressible component of the weighted velocity introduced in the main text, and 
$\boldsymbol{e}_r$ 
is the unit vector along $\boldsymbol{r}$. The brackets $\langle \cdot \rangle$ indicate averaging over all possible system positions $\boldsymbol{R}$ and orientations of 
$\boldsymbol{e}_r$. 
In the IR, Kolmogorov's 1941 (K41) theory predicts the power-law
\(
S_p(r) \propto r^{
\zeta_0 \left( p \right)
}
\), with $
\zeta_0 \left( p \right)
=p/3
$. For \(p = 2\), this scaling relation yields the incompressible kinetic energy spectrum \(E_i(k) \propto k^{-5/3}\) \cite{Kolmogorov41,Benzi1993,Dubrulle1994,Frisch1995}.
In Fig. \ref{fig:ESS_Tq_350} (a), we show the structure functions $S_{p}(r)$ up to order $p=5$. The Kolmogorov power-law regime appears narrow, making it difficult to reliably extract the scaling exponents. To overcome this, we additionally plot $S_{p}$ as a function of $S_{3}$ in Fig. \ref{fig:ESS_Tq_350} (c)-(f). Since $S_{3}\propto r$ according to K41, this representation preserves the same scaling exponents as $S_{p}(r)$.
Moreover, the scaling law extends beyond the IR, revealing the self-similar character of the velocity structure functions, known as extended self-similarity (ESS) \cite{Benzi1993}. 
In particular, Fig.~\ref{fig:ESS_Tq_350} demonstrates that the relation $S_p(r) \propto [S_3(r)]^{\zeta(p)}$ holds well beyond the inertial range for $\tau_Q = 350$, namely over $0.03 \lesssim r \lesssim l_v = 4.295$, while the corresponding inertial range lies within $d_v = 0.632 \lesssim r \lesssim l_v$ (see Fig.~2(b) in the main text). The fitted scaling relations in Figs.~\ref{fig:ESS_Tq_350} (c - f) are given by
$S_1(r) = \left( 0.420 \pm 0.011 \right) \left\lbrack S_3(r) \right\rbrack^{0.439 \pm 0.011}$, 
$S_2(r) = \left( 0.457 \pm 0.021 \right) \left\lbrack S_3(r) \right\rbrack^{0.820 \pm 0.022}$, 
$S_4(r) = \left( 3.316 \pm 0.220 \right) \left\lbrack S_3(r) \right\rbrack^{1.069 \pm 0.033}$, 
and
$S_5(r) = \left( 13.794 \pm 1.061 \right) \left\lbrack S_3(r) \right\rbrack^{1.118 \pm 0.038}$, 
obtained from the orthogonal distance regression 
using the whole range of $r$ in Fig. \ref{fig:ESS_Tq_350} (a), 
with the uncertainties in $S_p(r)$ given by their corresponding standard deviations. 
This broader scaling range allows for a more reliable determination of the Kolmogorov exponents, as shown in Fig. \ref{fig:ESS_Tq_350} (b). We observe deviations from the K41 prediction, $\zeta(p) = p/3$, 
consistent with results reported in both classical and quantum turbulence \cite{Benzi1993,Krstulovic2016,Zhao25}. Such deviations can be accounted for by a refinement of the K41 prediction. In particular, we test the K62 formulation, which introduces intermittency corrections and predicts $\zeta(p) = p / 3 - \kappa p (p - 3) / 2$,
where $\kappa$ is a constant \cite{Kolmogorov_1962}.
As shown in Fig. \ref{fig:ESS_Tq_350} (b), the measured exponents are in good agreement with the K62 prediction, with fitting parameter $\kappa = 0.114 \pm 0.006$.

\section{Universality of the compressible kinetic energy and its spectrum}
\begin{figure}[pb]
\includegraphics[width=0.82\columnwidth]{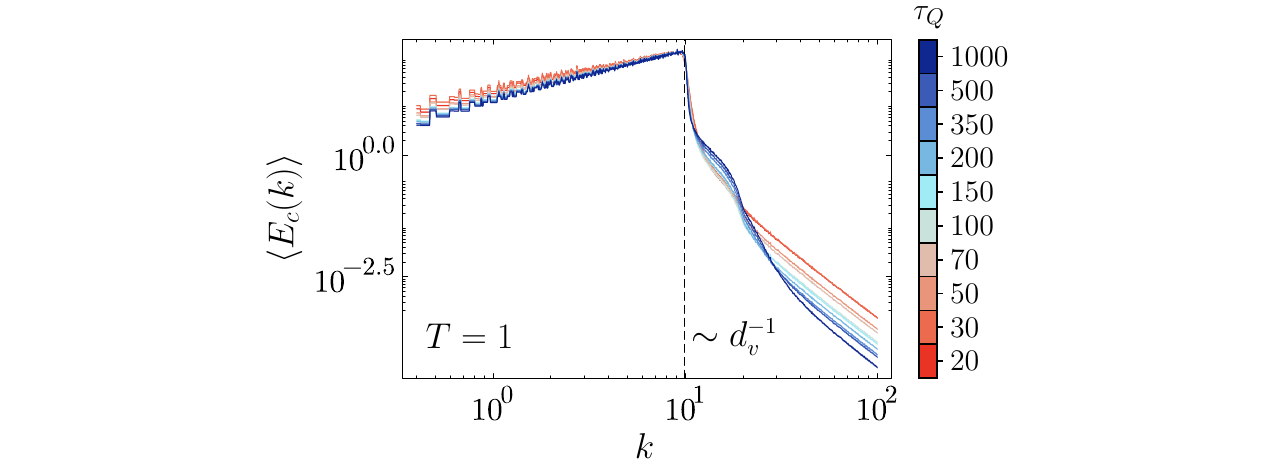}
\caption{Compressible kinetic energy spectra at temperature $T = 1$. 
Ensemble-averaged compressible kinetic energy spectra $\langle E_c(k)\rangle$ at equilibration time for various quench durations $\tau_Q$ at $T=1$. Each curve is obtained by averaging over $\mathcal{R}=1000$ stochastic realizations. The dashed vertical line indicates $k = 2\pi/d_v$, where $d_v = 4\xi_h$ estimates the vortex diameter, with $\xi_h$ the healing length. Shaded regions denote 95\% confidence intervals.
}
\label{fig:Ec_Tq_KZM_T_1}
\end{figure}

Having established the universal scaling of the incompressible kinetic energy and its spectrum, we now turn to analyze the compressible component across different quench durations $\tau_{Q}$.
Fig. \ref{fig:Ec_Tq_KZM_T_1} shows the corresponding spectra $E_{c}(k)$, which exhibit a substantial collapse onto a single curve, indicating an approximate independence from $\tau_{Q}$. 
\begin{figure}[hptb]
\includegraphics[width=0.85\columnwidth]{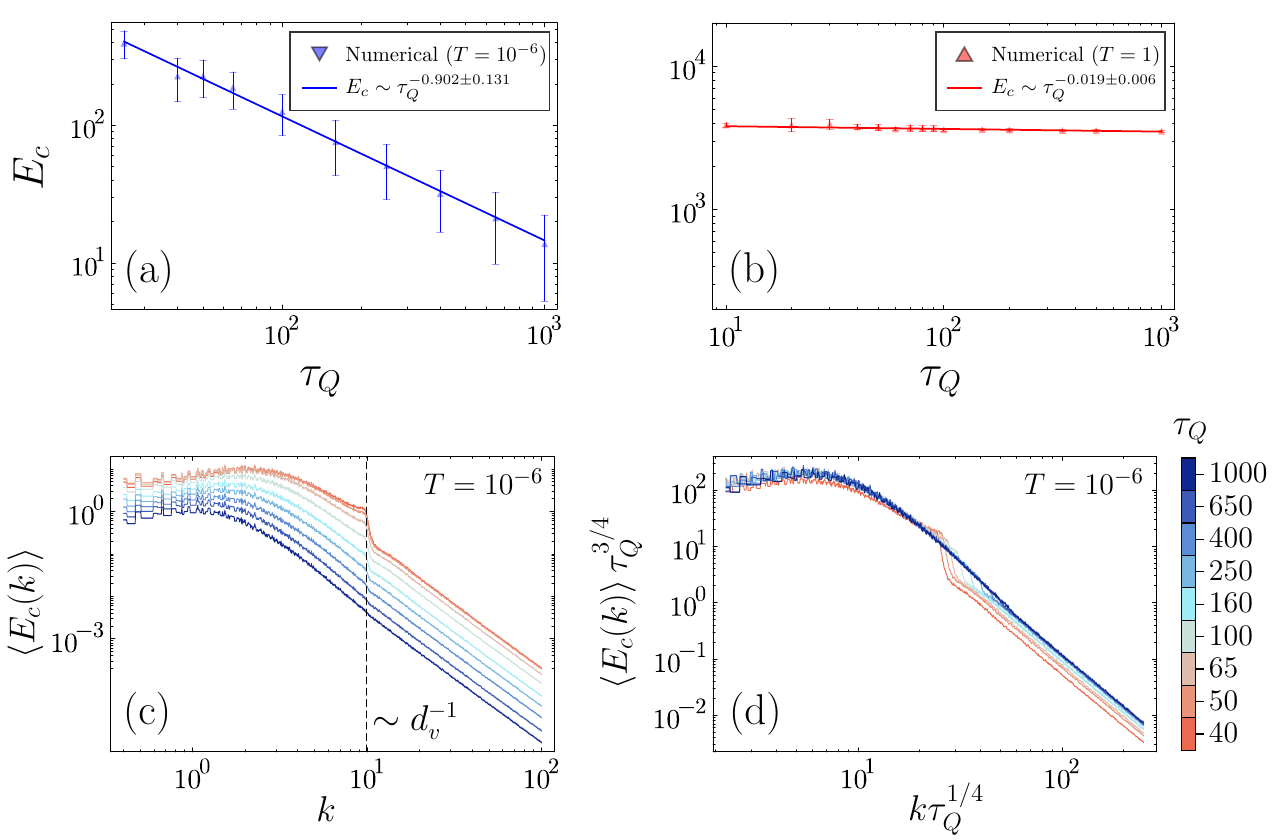}
\caption{Kibble–Zurek universality of the compressible kinetic energy. 
Panels (a) and (b) show the compressible kinetic energy $E_c$ at equilibration for various quench durations $\tau_Q$ at temperatures $T = 10^{-6}$ and $T = 1$, respectively. Each curve in (a) and (b) is obtained by averaging over $\mathcal{R}=2000$ and $\mathcal{R}=1000$ stochastic realizations, respectively. Error bars denote one standard deviation.
Panel (c) shows the ensemble-averaged spectrum $\langle E_c(k) \rangle$ at equilibration for several values of $\tau_Q$ at $T=10^{-6}$, each curve averaged over $\mathcal{R}=2000$ realizations. The dashed vertical line indicates $k = 2\pi/d_v$, where $d_v = 4\xi_h$ estimates the vortex diameter, with $\xi_h$ the healing length.
Panel (d) displays $\langle E_c(k) \rangle \tau_Q^{3/4}$ as a function of the scaled momentum $k \tau_Q^{1/4} \propto k\hat{\xi}$. In panels (c) and (d), shaded error bands denote 95\% confidence intervals.
}
\label{fig:Ec_integ_Tq_KZM}
\end{figure}
This is also reflected in the total compressible energy $E_{c}$, shown in Fig.~\ref{fig:Ec_integ_Tq_KZM} (b), where a power-law fit yields $E_{c} = (3.977 \pm 0.120)\times 10^{3}\,\tau_{Q}^{-0.019 \pm 0.006}$.
The reason for this behavior is that, at the simulation temperature $T=1$ (corresponding to a physical temperature of $O(1)\, \text{nK}$ for parameters relevant to homogeneous quasi-2D BEC experiments \cite{Chomaz15} and used throughout the main text), condensate-density modulations contributing to $E_{c}$ are predominantly due to thermal fluctuations, and are therefore independent of $\tau_Q$. This interpretation is supported by additional simulations at $T=10^{-6}$ ($O(10^{-6})\,\text{nK}$), where the magnitude of $E_{c}$ decreases by nearly two orders of magnitude compared to the $T=1$ case (see Fig. \ref{fig:Ec_integ_Tq_KZM} (a,b)).

At $T=10^{-6}$, the suppression of thermal fluctuations allows density modulations arising from phonons emitted during vortex–antivortex annihilation 
\cite{Kumar2025} 
to contribute more noticeably to the compressible energy. The corresponding spectra for different $\tau_{Q}$ are shown in Fig. \ref{fig:Ec_integ_Tq_KZM} (c), and, unlike the $T=1$ case, they are clearly distinct. This naturally raises the question of whether a rescaling in $\tau_{Q}$ can restore their collapse.
To address this, we assume that $E_c$ originates purely from vortex–antivortex annihilations.
The total number of emitted phonons at equilibration, $N_p$, is proportional to the number of annihilation events and therefore to the total number of vortices $n_v$ in a periodic system, where vortices and antivortices appear in equal numbers.
According to the Kibble--Zurek mechanism (KZM), this implies $N_p \propto n_v \propto \hat{\xi}^{-2} \propto \tau_Q^{-2\nu/(1 + z \nu)}$ in two spatial dimensions, where $\hat{\xi}$ is the Kibble-Zurek correlation length and $\nu$ and $z$ are the correlation-length and dynamic critical exponents, respectively. Let $n_p(k)$ denote the phonon number density at momentum $
\boldsymbol{k}
$, defined such that $\int_0^\infty n_p(k)\,dk = N_p$. At equilibration, $\hat{\xi}$ sets the characteristic momentum scale of the spectrum, motivating the scaling ansatz $n_p(k) = \tau_Q^{-\alpha} F_p(k\hat{\xi})$, where $F_p(x)$ is a function of the scaling variable $x = k\hat{\xi}$.
Integrating over all momenta gives
\begin{equation}
\int_0^\infty n_p(k)\,dk \propto \tau_Q^{-\alpha} \hat{\xi}^{-1}
\propto \tau_Q^{-\alpha - \nu/(1 + z \nu)},
\end{equation}
which fixes $\alpha = \nu/(1 + z \nu)$. 
Each phonon carries energy $
k
c_s
$, where $
c_s = 
\sqrt{
\mu}
$ is the speed of sound in the condensate \cite{pitaevskii2003bose}. 
For a linear quench of the chemical potential $\mu(t)$, we have $\mu(t_{\textrm{eq}}) \propto \tau_Q^{-1/(1 + z \nu)}$, giving the total phonon energy
\begin{equation}\label{E_p_tot_eq}
E_p^{\mathrm{tot}}
= \int_0^\infty n_p(k)\,
k 
c_s
\,dk
\propto
\tau_Q^{-\alpha}
\tau_Q^{-2\nu/(1 + z \nu)}
\sqrt{\mu(t_{\textrm{eq}})}
\propto
\tau_Q^{-(3\nu + 1/2)/(1 + z \nu)}.
\end{equation}
The corresponding spectral density is
\begin{equation}\label{E_p_k_eq}
E^{\textrm{tot}}_p (k) 
\propto 
\tau_Q^{- \left( 2 \nu + 1 / 2 \right) / \left( 1 + z \nu \right)} 
G \left( k \tau_Q^{\nu / \left( 1 + z \nu \right)} \right),
\end{equation}
where $G$ denotes the general function encoding the spectral shape.
For the mean-field exponents $\nu = 1/2$ and $z = 2$, Eq. (\ref{E_p_tot_eq}) gives $E_p^{\mathrm{tot}} \propto \tau_Q^{-1}$. 
Hence, if $E_c$ is primarily due to vortex–antivortex annihilation, one expects $E_c \propto \tau_Q^{-1}$. 
This prediction is tested in Fig.~\ref{fig:Ec_integ_Tq_KZM} (a), where the fitted power law is 
$E_c = (7.422 \pm 4.679)\times 10^3\,\tau_Q^{-0.902 \pm 0.131}$.
Furthermore, Eq. (\ref{E_p_k_eq}) implies that $\ E_c(k)\tau_Q^{3/4}$ should collapse when plotted against $k\tau_Q^{1/4}$. Such a collapse is indeed observed in Fig.~\ref{fig:Ec_integ_Tq_KZM} (d).
As a final remark, we note that although the rescaling in $\tau_{Q}$ required to obtain an overlap of the spectra is the same in the mean-field regime for both the compressible and incompressible energy spectra, this is purely a coincidence due to the mean-field values of the critical exponents. In fact, the corresponding general scaling forms differ (compare Eq. (\ref{E_p_k_eq}) with Eq. (4) of the main text).
\bigskip

\section{Time evolution of the incompressible kinetic energy spectrum, enstrophy, and vortex number}
\begin{figure}[pb]
    \centering
    \includegraphics[width=0.94\linewidth]{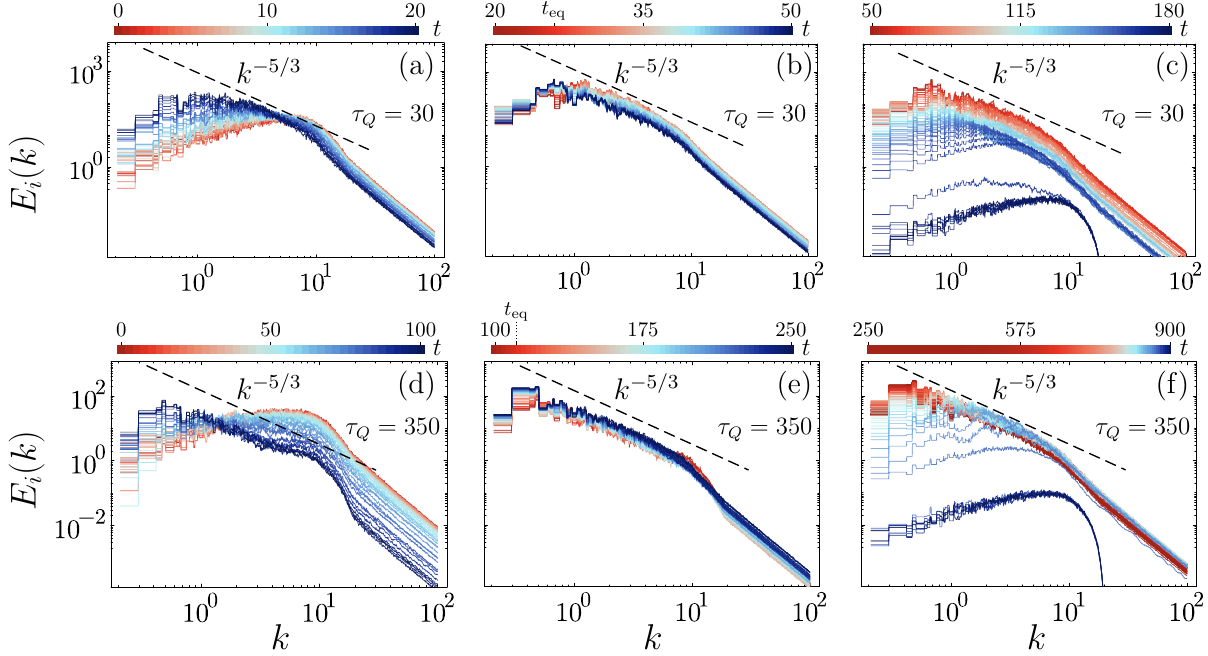}
    \caption{Time evolution of the incompressible kinetic energy spectrum. Panels (a), (b) and (c) show subsequent intervals of time evolution of the incompressible kinetic energy spectrum $E_{i}(k)$ in a single trajectory of the system for a quench of duration $\tau_{Q}=30$. The equilibration time in this case is $t_{\textrm{eq}}=25.9$. Panels (d), (e) and (f) show the same for $\tau_{Q}=350$, where $t_{\textrm{eq}}=111.5$. In both cases the quench starts at $t=0$.}
    \label{fig:E_i(k)_time_evolution_Tq_30_350}
\end{figure}

Throughout this work, we have focused on the study of spontaneous quantum turbulence at equilibration time, when the predictions of the Kibble-Zurek (KZ) mechanism are valid. It is however also interesting to consider the full time evolution of the system to observe how the turbulent dynamics unfold. Specifically, in Fig. \ref{fig:E_i(k)_time_evolution_Tq_30_350}, we show the time evolution of the incompressible kinetic energy spectrum $E_{i}(k)$ in a single simulation run for a fast quench ($\tau_{Q}=30$, upper panels) and a slower quench ($\tau_{Q}=350$, lower panels). Near the equilibration time, the $k^{-5/3}$ Kolmogorov scaling emerges, signaling the onset of spontaneous quantum turbulence in the newborn condensate. This scaling regime remains visible during the subsequent evolution until all vortices are annihilated, as indicated by a sudden decay in the amplitude of $E_{i}(k)$.
The extent of the inertial range, where the $k^{-5/3}$ power law is observed, is relatively narrow for the fast quench ($\tau_{Q}=30$, see Fig.~\ref{fig:E_i(k)_time_evolution_Tq_30_350}(b)), whereas for the slower quench ($\tau_{Q}=350$, see Fig.~\ref{fig:E_i(k)_time_evolution_Tq_30_350}(e)) it becomes broader, spanning nearly a decade in $k$. This behavior is consistent with the range being bounded by the dissipation scale, set by the vortex-core size, and the injection scale, given by the inter-vortex distance determined by the KZ length $\hat{\xi}$: as $\hat{\xi}$ decreases for faster quenches, the inertial-range window naturally shrinks.

\begin{figure}[pt]
\centering
  \includegraphics[width=0.42\linewidth]{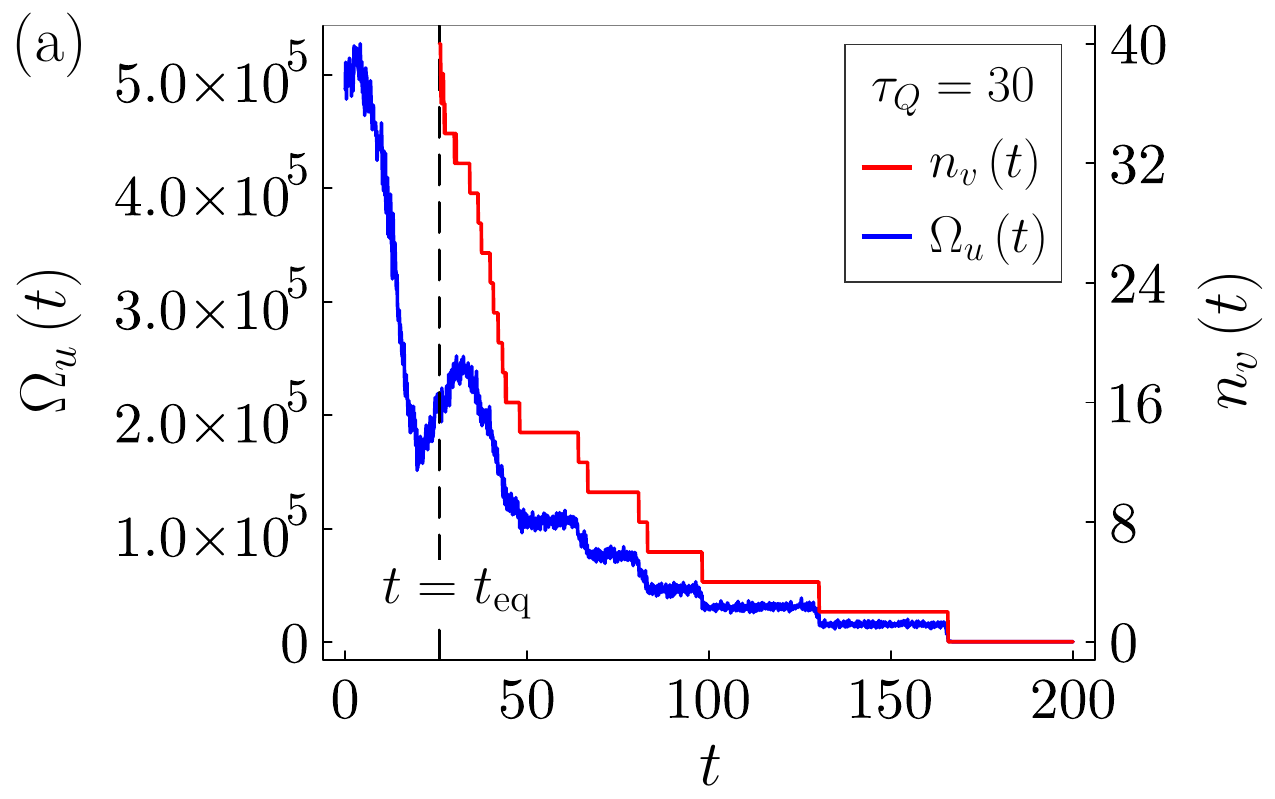}
  \qquad 
  \includegraphics[width=0.42\linewidth]{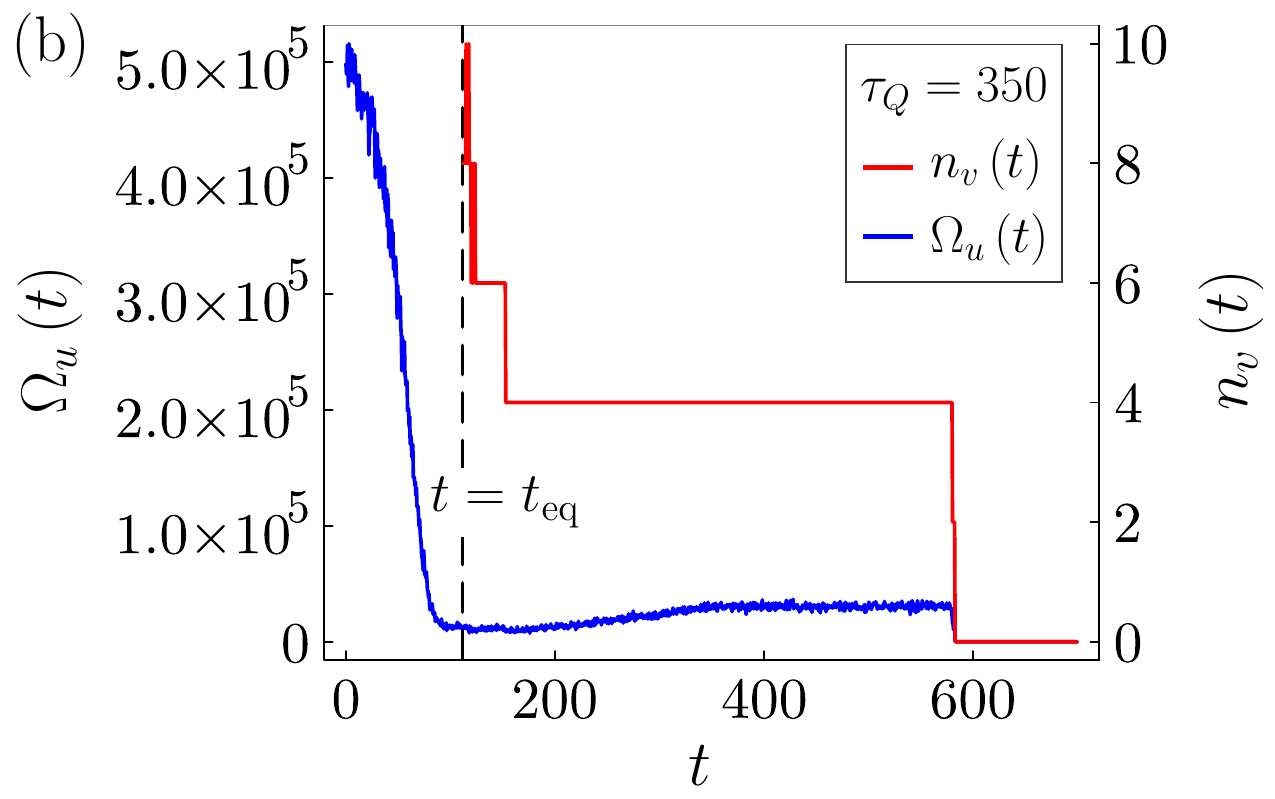}
  \caption{Number of vortices $n_v \left( t \right)$ and enstrophy $\Omega_u \left( t \right) = \int \left\vert \nabla \times \boldsymbol{u} \left( \boldsymbol{r}, t \right) \right\vert^2 \; d^2 \boldsymbol{r}$ as functions of time $t$. The data are from the same $T=1$ simulations as in Fig. \ref{fig:E_i(k)_time_evolution_Tq_30_350}, with quench times (a) $\tau_Q=30$ ($t_{\textrm{eq}}=25.9$) and (b) $\tau_Q=350$ ($t_{\textrm{eq}}=111.5$).
}
  \label{fig:enstrophy_vs_time}
\end{figure}
Besides the energy spectrum, we have also examined the time evolution of the enstrophy, defined as
\begin{equation}
\Omega_u(t) = \int \left| \nabla \times \boldsymbol{u}(\boldsymbol{r}, t) \right|^2 \, d^2\boldsymbol{r},
\end{equation}
where $\nabla \times \boldsymbol{u}$ denotes the vorticity associated with the weighted superfluid velocity field 
(we considered the curl of the weighted velocity $\boldsymbol{u}$, since the integration of the curl of the velocity is proportional to the number of vortices).
In two-dimensional (2D) classical turbulence, the conservation of enstrophy plays a key role, as it leads to an inverse energy cascade characterized by the formation of increasingly larger eddies.
In quantum turbulence (QT), a similar mechanism can occur through the clustering of same-sign vortices, resulting in extended regions of high vorticity. However, in 2D 
QT, 
enstrophy is not necessarily conserved, and this can give rise to a direct energy cascade \cite{Numasato2010,Chesler13,Zhao25}.
Fig. 
\ref{fig:enstrophy_vs_time} shows the time evolution of the enstrophy together with the vortex count from the same data used in Fig. \ref{fig:E_i(k)_time_evolution_Tq_30_350}, for $\tau_{Q}=30$ in panel (a) and $\tau_{Q}=350$ in (b). The results indicate that enstrophy is not conserved throughout the evolution, suggesting the presence of a direct energy cascade. Moreover, tracking the vortex positions over time reveals 
no formation of same-sign vortex clusters.

\section{Numerical evaluation of the compressible and incompressible energy spectra}
Given the weighted superfluid velocity $\boldsymbol{u}$, we outline the procedure to extract its irrotational and solenoidal components, $\boldsymbol{u}_{c}$ and $\boldsymbol{u}_{i}$, respectively.

Using the Green's function \cite{Watanabe2013}, any 
two-dimensional vector field $\boldsymbol{V} \left( \boldsymbol{r}\right)$ can be expressed as 
\begin{equation}\label{vector_integral_identity}
\boldsymbol{V} \left( \boldsymbol{r}\right) 
= 
\int 
\boldsymbol{V} \left( \boldsymbol{r}^{\prime}\right) 
\delta \left( \boldsymbol{r} - \boldsymbol{r}^{\prime} \right) 
\, d^2 \boldsymbol{r}^{\prime} 
= 
\nabla^2 
\left\lbrack 
\frac{1}{2 \pi} 
\int 
\boldsymbol{V} \left( \boldsymbol{r}^{\prime}\right) 
\ln \left( \frac{\left\vert \boldsymbol{r} - \boldsymbol{r}^{\prime} \right\vert}{l} \right) 
\, 
d^2 \boldsymbol{r}^{\prime} 
\right\rbrack 
, 
\end{equation}
where $\delta \left( \boldsymbol{r} \right)$ is the two-dimensional Dirac delta function 
and $l$ is a constant with units of length. 
Applying Eq. (\ref{vector_integral_identity}) to $\boldsymbol{u}(\boldsymbol{r})$ yields $\boldsymbol{u}(\boldsymbol{r})=\nabla^{2}\boldsymbol{b}(\boldsymbol{r})$, where $
\boldsymbol{b} \left( \boldsymbol{r}\right) 
\coloneqq 
\left( 1 / 2 \pi \right) 
\int 
\boldsymbol{u} \left( \boldsymbol{r}^{\prime} \right) 
\ln \left( \left\vert \boldsymbol{r} - \boldsymbol{r}^{\prime} \right\vert / l \right) 
\, d^2 \boldsymbol{r}^{\prime} 
$.
Using the vector identity $\nabla \times \left( \nabla \times \boldsymbol{A} \right) = \nabla \left( \nabla \cdot \boldsymbol{A} \right) - \nabla^2 \boldsymbol{A}$, we can thus decompose $\boldsymbol{u}$ into:
\begin{equation}
\boldsymbol{u}_{c} \left( \boldsymbol{r}, t \right) 
= 
\nabla 
\left\lbrack 
\nabla 
\cdot 
\boldsymbol{b} \left( \boldsymbol{r}, t \right) 
\right\rbrack 
, \quad 
\boldsymbol{u}_{i} \left( \boldsymbol{r}, t \right) 
= 
- 
\nabla 
\times 
\left\lbrack 
\nabla 
\times 
\boldsymbol{b} \left( \boldsymbol{r}, t \right) 
\right\rbrack,
\label{Eq:app:B:vc_vic_as_b}
\end{equation}
satisfying 
$\nabla \times \boldsymbol{u}_{c} \left( \boldsymbol{r}, t \right) = 0$ and 
$\nabla \cdot \boldsymbol{u}_{i} \left( \boldsymbol{r}, t \right) = 0$.

In Fourier space, we have $\boldsymbol{\tilde{u}}(\boldsymbol{k})=-{k}^{2}\boldsymbol{\tilde{b}}(\boldsymbol{k})$, with $\boldsymbol{\tilde{b}} \left( \boldsymbol{k}, t \right) 
= 
\left( 1 / 2 \pi \right) 
\int 
e^{- i \boldsymbol{k} \cdot \boldsymbol{r}} 
\boldsymbol{b} \left( \boldsymbol{r}, t \right) 
\, d^2 \boldsymbol{r} $.
Inserting this into Eq. (\ref{Eq:app:B:vc_vic_as_b}) gives:
\begin{equation}\label{u_c_and_u_i_expressed_as_k_integrals}
\boldsymbol{u}_{c} \left( \boldsymbol{r}, t \right) 
=  
\frac{1}{2 \pi} 
\int 
\frac{\boldsymbol{k}}{k^2} 
\left\lbrack 
\boldsymbol{k} 
\cdot 
\boldsymbol{\tilde{u}} \left( \boldsymbol{k}, t \right) 
\right\rbrack 
e^{i \boldsymbol{k} \cdot \boldsymbol{r}} 
\, d^2 \boldsymbol{k} 
, \quad 
\boldsymbol{u}_{i} \left( \boldsymbol{r}, t \right) 
= 
- 
\frac{1}{2 \pi} 
\int 
\frac{\boldsymbol{k}}{k^2} 
\times 
\left\lbrack 
\boldsymbol{k} 
\times 
\boldsymbol{\tilde{u}} \left( \boldsymbol{k}, t \right) 
\right\rbrack 
e^{i \boldsymbol{k} \cdot \boldsymbol{r}} 
\, d^2 \boldsymbol{k} 
, 
\end{equation}
where $k \coloneqq \left\vert \boldsymbol{k} \right\vert$. 
Thus, in momentum space, the compressible and incompressible components of $\boldsymbol{u}$ read
$
\boldsymbol{\tilde{u}}_{c} \left( \boldsymbol{k}, t \right) 
= 
\boldsymbol{e}_k 
\left\lbrack 
\boldsymbol{e}_k 
\cdot 
\boldsymbol{\tilde{u}} \left( \boldsymbol{k}, t \right) 
\right\rbrack 
$, and 
$
\boldsymbol{\tilde{u}}_{i} \left( \boldsymbol{k}, t \right) 
= 
\boldsymbol{\tilde{u}} \left( \boldsymbol{k}, t \right) 
- 
\boldsymbol{\tilde{u}}_{c} \left( \boldsymbol{k}, t \right) 
$ where $\boldsymbol{e}_k \coloneqq \boldsymbol{k} / k$.
Finally, the compressible and incompressible energy spectra are given by
\begin{equation}\label{E_c_k_and_E_i_k_equations}
E_{c} \left( k 
, t 
\right) 
= 
\frac{k}{2}
\int_{0}^{2 \pi} 
\left\lbrack 
\boldsymbol{e}_k 
\cdot 
\boldsymbol{\tilde{u}}^{*} \left( \boldsymbol{k}, t \right) 
\right\rbrack 
\left\lbrack 
\boldsymbol{e}_k 
\cdot 
\boldsymbol{\tilde{u}} \left( \boldsymbol{k}, t \right) 
\right\rbrack 
\, d \varphi_k 
, \quad 
E_{i} \left( k 
, t 
\right) 
= 
\frac{k}{2}
\int_{0}^{2 \pi} 
\boldsymbol{\tilde{u}}^{*} \left( \boldsymbol{k}, t \right) 
\cdot 
\boldsymbol{\tilde{u}} \left( \boldsymbol{k}, t \right) 
\, d \varphi_k 
- 
E_{c} \left( k \right) 
, 
\end{equation}
where $\varphi_k$ is the azimuthal angle of $\boldsymbol{k}$, i.e., $\boldsymbol{e}_k = \left( \cos \varphi_k, \sin \varphi_k \right)$. 
As our analysis mainly focuses on the system at equilibration time $t_{\textrm{eq}}$, we will omit the time argument for brevity and denote $E_{c, i}(k) \coloneqq E_{c, i}(k, t_{\textrm{eq}})$. 

\begin{figure}[hptb]
\centering
\includegraphics[width=0.8 \linewidth]{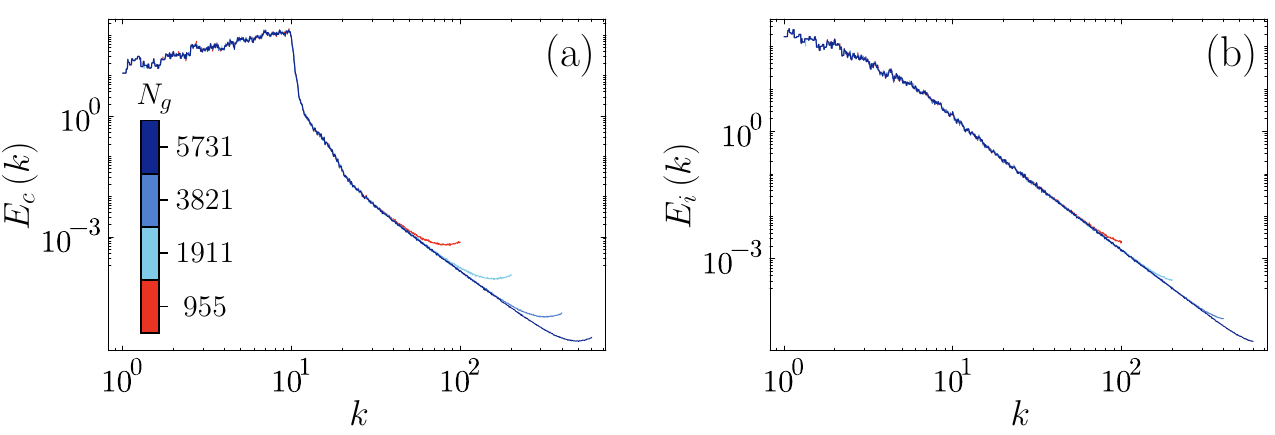}
\caption{
\textbf{Kinetic energy spectrum calculated using the quadrature 
method}. 
(a) Compressible kinetic energy spectrum $E_c(k)$ as a function of $k$ for different grid sizes $N_g$. 
(b) Incompressible kinetic energy spectrum $E_i(k)$ as a function of $k$ for the same $N_g$. 
Both panels correspond to a quench time of $\tau_Q = 60$, evaluated at the equilibration time. 
For this analysis, a single realization was selected from the $1000$ independent noise realizations to determine the optimal $N_g$. 
Each $N_g$ is chosen according to $N_g = 2 \lfloor L k_m / 2 \pi \rceil + 1$, with $k_m = 100$, $200$, $400$, and $600$.
}
\label{fig:Ec_Eic_quadrature}
\end{figure}

To numerically evaluate $E_{c,i}(k)$ as given in 
Eqs. (\ref{E_c_k_and_E_i_k_equations}),
we used the Julia package \texttt{FFTW.jl}, which provides an interface to the \texttt{FFTW3} library \cite{FFTW05}, allowing efficient computation of the fast Fourier transform to obtain  $\boldsymbol{\tilde{u}} \left( \boldsymbol{k}\right)$. 
To perform the angular integral over $\varphi_k$, we employed the quadrature method introduced in \cite{blakie2008dynamics}. 
Denoting by $k_m$ the maximum wavenumber considered for evaluating 
$E_{c,i}(k)$, 
and by $N_g$ the number of grid points along each spatial direction for a periodic system of size $L \times L$, we compute and plot $E_c(k)$ and $E_i(k)$ for different values of $N_g$ in Fig. \ref{fig:Ec_Eic_quadrature}. 
It 
clearly shows that sufficiently large values of $N_g$ are required to suppress numerical errors at high wavenumbers $k$. We found that the minimal grid size needed to ensure accuracy for $k \leq k_m$ is approximately $N_g \approx 8 \lfloor L k_m / 2 \pi \rceil + 1$, where $\lfloor x \rceil$ denotes the nearest integer to $x$. Accordingly, we used $N_g = 3817$ to accurately compute 
$E_{c,i}(k)$ 
up to $k \leq 100$.

\section{Numerical evaluation of the velocity structure function}
To numerically calculate the velocity structure function 
\begin{equation}
S_p \left( r \right) 
\coloneqq 
\frac{1}{2 \pi L^2} 
\int_{0}^{2 \pi} 
\int_{0}^{L} 
\int_{0}^{L} 
\left\vert 
\left\lbrack 
\boldsymbol{u}_i \left( \boldsymbol{R} + \boldsymbol{r} \right) 
- 
\boldsymbol{u}_i \left( \boldsymbol{R} \right) 
\right\rbrack 
\cdot 
\boldsymbol{e}_r 
\right\vert^p 
\; 
d R_x \; 
d R_y \; 
d \varphi 
, 
\label{supp:eq:Sp}
\end{equation}
where $\boldsymbol{e}_r \coloneqq \boldsymbol{r} / r = \left( \cos \varphi, \sin \varphi \right)$ is the unit vector along $\boldsymbol{r}$ \cite{Ciliberto1994,Frisch1995}, we compute the velocity increments using the discrete Fourier transform, implemented via the Julia package \texttt{FFTW.jl}. 
Let $\nu_j \coloneqq \left( j - 1 \right) L_{\nu} / N_{\nu}$ for $1 \le j \le N_{\nu}$ with $\nu = x, y$. The weighted superfluid velocity $\boldsymbol{u} \left( x_j, y_l\right)$ can then be written as 
\begin{equation}
\boldsymbol{u} \left( x_j, y_l \right) 
= 
\frac{1}{N_x N_y} 
\sum_{n_x = 1}^{N_x} 
\sum_{n_y = 1}^{N_y} 
\boldsymbol{\tilde{u}}\left( n_x - 1, n_y - 1 \right) 
e^{2 \pi i \left( n_x - 1 \right) \left( j - 1 \right) / N_x}
e^{2 \pi i \left( n_y - 1 \right) \left( l - 1 \right) / N_y}
,
\end{equation}
where we used $\boldsymbol{k} / 2 \pi = \left( \left( n_x - 1 \right) / L_x, \left( n_y - 1 \right) / L_y \right)$ in the discrete Fourier space. 
Since $\boldsymbol{u} \left( x_j, y_l \right)$ is real, its derivatives must also be real. 
Using the identity $\exp \left( 2 \pi i \left( n_{\nu} - 1 \right) \left( j - 1 \right) / N_{\nu} \right) = \exp \left( 2 \pi i \left( n_{\nu} - 1 - N_{\nu} \right) \left( j - 1 \right) / N_{\nu} \right)$ for any integers $1 \le n_{\nu} \le N_{\nu}$ and $1 \le j \le N_{\nu}$, in order to make $\boldsymbol{u}_i \left( x_j, y_l \right)$ real, one should use the following expression: 
\begin{equation}
\boldsymbol{u}_i \left( x_j, y_l \right) 
= 
\frac{1}{N_x N_y} 
\sum_{n_x = 1}^{N_x} 
\sum_{n_y = 1}^{N_y} 
\frac{
\boldsymbol{k} \left( n_x, n_y \right) 
}{
k^2 \left( n_x, n_y \right) 
} 
\times 
\left\lbrack 
\boldsymbol{\tilde{u}}\left( n_x - 1, n_y - 1 \right) 
\times 
\boldsymbol{k} \left( n_x, n_y \right) 
\right\rbrack 
e^{2 \pi i \left( n_x - 1 \right) \left( j - 1 \right) / N_x}
e^{2 \pi i \left( n_y - 1 \right) \left( l - 1 \right) / N_y}
, 
\label{eq:app:u_i_discrete_space}
\end{equation}
with $\boldsymbol{k} \left( n_x, n_y \right) \coloneqq \sum_{\nu = x, y} k_{\nu} \left( n_x, n_y \right) \boldsymbol{e}_{\nu} $ and 
\begin{equation}
k_{\nu} \left( n_x, n_y \right) 
\coloneqq 
\left\{ 
\begin{array}{cc}
2 \pi i \left( n_{\nu} - 1 \right) / L_{\nu} 
, 
& 
\textrm{ if \; $1 \le n_{\nu} \le \left\lbrack N_{\nu} / 2 \right\rbrack_r + 1$,}
\\
\\
2 \pi i \left( n_{\nu} - 1 - N_{\nu} \right) / L_{\nu} 
, 
& 
\textrm{ otherwise,}
\end{array}
\right. 
\end{equation}
where $\left\lbrack x \right\rbrack_r$ is the greatest integer less than or equal to $x$, so that $- 2 \pi i \left\lbrack N_{\nu} / 2 \right\rbrack_r / L_{\nu} \le k_{\nu} \left( n_x, n_y \right) \le 2 \pi i \left\lbrack N_{\nu} / 2 \right\rbrack_r / L_{\nu}$.

Note that, for any function $f \left( x, y \right)$ in the periodic system $f \left( x + L_x, y \right) = f \left( x, y + L_y \right) = f \left( x, y \right)$, 
\begin{equation}
\frac{1}{L_x L_y} 
\int_0^{L_x} 
\int_0^{L_y} 
f \left( x, y \right) \; 
d y \; 
d x 
= 
\frac{1}{N_x N_y} 
\sum_{j = 1}^{N_x} 
\sum_{l = 1}^{N_y} 
f \left( x_j, y_l \right) 
. 
\label{eq:app:spatial_numerical_integration}
\end{equation}
After numerically calculating $\boldsymbol{u}_i \left( x_j, y_l \right)$ via Eq. \eqref{eq:app:u_i_discrete_space}, 
we calculate 
$
\Delta \boldsymbol{u}_i \left( \boldsymbol{R}, \boldsymbol{r} \right) 
\coloneqq 
\boldsymbol{u}_i \left( \boldsymbol{R} + \boldsymbol{r} \right) 
- 
\boldsymbol{u}_i \left( \boldsymbol{R} \right) 
$ and then calculate $S_p \left( r \right)$ numerically as 
\begin{equation}
S_p \left( r \right) 
= 
\frac{1}{L^2} 
\int_{0}^{L} 
\int_{0}^{L} 
\left\lbrack 
\frac{1}{2 \pi} 
\int_0^{2 \pi} 
\left\vert 
\Delta \boldsymbol{u}_i \left( \boldsymbol{R}, \boldsymbol{r} \right) 
\cdot 
\boldsymbol{e}_r 
\right\vert^p \; 
d \varphi 
\right\rbrack \;
d R_x \; 
d R_y 
= 
\frac{1}{N^2_x} 
\sum_{j, l = 1}^{N_x} 
\left\lbrack 
\frac{1}{2 \pi} 
\int_{0}^{2 \pi} 
\left\vert 
\Delta \boldsymbol{u}_i \left( X_j, Y_l, \boldsymbol{r} \right) 
\cdot 
\boldsymbol{e}_r 
\right\vert^p \; 
d \varphi 
\right\rbrack 
,
\end{equation}
with $X_j \coloneqq \left( j - 1 \right) L / N_x$ and $Y_l \coloneqq \left( l - 1 \right) L / N_x$.

\end{document}